\documentclass[journal]{IEEEtran}
\usepackage{setspace}
\usepackage{cite}
\ifCLASSINFOpdf
   \usepackage[pdftex]{graphicx}
\else
\fi
\usepackage[cmex10]{amsmath}
\usepackage{amsmath, amssymb}
\usepackage{amsthm}
\usepackage{algorithmic}
\usepackage{algorithm}

\usepackage{array}
\usepackage[tight,footnotesize]{subfigure}
\usepackage{graphicx}
\usepackage{caption}
\begin{document}
%
% paper title
% can use linebreaks \\ within to get better formatting as desireds
\title{Massive MIMO as a Big Data System: Random Matrix Models and Testbed}
%\title{ Initial Experimental Demonstration of Cooperative Spectrum Sensing over Large Scale Software Defined Radio Network Testbed using Random Matrix Modeling}

%\author{\IEEEauthorblockN{Changchun Zhang, Zhen Hu, Robert C. Qiu}
%\IEEEauthorblockA{Cognitive Radio Insitute,
%Department of Electrical and Computer Engineering, \\Center for Manufacturing Research, Tennessee Technological University, \\Cookeville, Tennessee, USA\\
%Email: \{czhang42\}@students.tntech.edu, \{rqiu\}@tntech.edu}}

 %\author{\IEEEauthorblockN{Changchun Zhang and  Robert C. Qiu
 %}
%\IEEEauthorblockA{Cognitive Radio Insitute,
%Department of Electrical and Computer Engineering, \\Center for Manufacturing Research, Tennessee Technological University, \\Cookeville, Tennessee 38505, USA\\
%Email: czhang42@students.tntech.edu, rqiu@tntech.edu}
%
%%\IEEEauthorblockA{\IEEEauthorrefmark{3}Sensor Systems Division, University of Dayton Research Institute\\
%%300 College Park, Dayton, OH 45469, USA\\Email: Michael.Wicks@udri.udayton.edu}
 %}

\author{        Changchun Zhang,~\IEEEmembership{Student Member,~IEEE}, Robert C. Qiu,~\IEEEmembership{Fellow,~IEEE}
       % <-this % stops a space
\thanks{Robert C. Qiu and Changchun Zhang are with the Department of Electrical and Computer Engineering, Center for Manufacturing Research, Tennessee Technological University, Cookeville, TN, 38505, e-mail: czhang42@students.tntech.edu; rqiu@ieee.org.}% <-this % stops a space
%\thanks{J. P. Browning is with Air Force Research Laboratory, Wright-Patterson AFB, Dayton, OH 45433, e-mail: James.Browning@wpafb.af.mil and M. C. Wicks is with Sensor Systems Division, University of Dayton Research Institute, Dayton, OH 45469, e-mail: Michael.Wicks@udri.udayton.edu.}% <-this % stops a space
%\thanks{Manuscript received April 19, 2005; revised January 11, 2007.}
}

\maketitle

\begin{abstract}
The paper has two parts. The first one deals with how to use large random matrices as building blocks to model the massive data arising from the massive (or large-scale) MIMO system. As a result, we apply this model for distributed spectrum sensing and network monitoring. The part boils down to the streaming, distributed massive data, for which a new algorithm is obtained and its performance is derived using the central limit theorem that is recently obtained in the literature. The second part deals with the large-scale testbed using software-defined radios (particularly USRP) that takes us more than four years to develop this 70-node network testbed. To demonstrate the power of the software defined radio, we reconfigure our testbed quickly into a testbed for massive MIMO. The massive data of this testbed is of central interest in this paper. It is for the first time for us to model the experimental data arising from this testbed. To our best knowledge, we are not aware of other similar work. 
\end{abstract}

\section{Introduction}

%Big data method demands 
%
%
%introduce the big data to the wireless communications.
%Data representation for wireless network  
%The cognitive radio network as sensors stores all the waveform data.
%
%Random matrix theory is the cornerstone of modeling our wireless big data.
%
%We show initial network status detection at waveform level, by analyzing the spectra of product of non-hermitian random matrices in a large cognitive radio networks.

Massive or large-scale multiple-input, multiple output (MIMO),  one of the disruptive technologies of the next generation (5G) communications system, promises significant gains in wireless data rates and link reliability ~\cite{rusek2013scaling} ~\cite{boccardi2013five}. In this paper, we deal with the massive data aspects of the massive MIMO system. In this paper, we use two terms (massive data and big data) interchangeably,  following the practice from National Research Council~\cite{frontiersmassive2013}. 

The benefits from massive MIMO are not only limited to the higher data rates. Massive MIMO techniques makes green communications possible. By using large numbers of antennas at the base station, massive MIMO helps to focus the radiated energy toward the intended direction while minimizing the intra and intercell interference. The energy efficiency is increased dramatically as the energy can be focused with extreme sharpness into small regions in space \cite{edfors2014massive}. It is shown in \cite{ngo2013energy} that, when the number of base station (BS) antennas $M$ grows without bound, we can reduce the transmitted power of each user proportionally to $1/M$ if the BS has perfect channel state information (CSI), and proportionally to $\frac{1}{\sqrt{M}}$ if CSI is estimated from uplink pilots. Reducing the transmit power of the mobile users can drain the batteries slower. Reducing the RF power of downlink can cut the electricity consumption of the base
station.

Massive MIMO also brings benefits including inexpensive low-power components, reduced latency, simplification of MAC layer, etc \cite{edfors2014massive}. Simpler network design could bring lower complexity computing which save more energy of the network to make the communications green.

Currently, most of the research of massive MIMO is focused on the communications capabilities. In this paper, we promote an insight that, very \textit{naturally}, the massive MIMO system can be regarded as a big data system. Massive waveform data---coming in a streaming manner---can be stored and processed at the base station with a large number of antennas, while not impacting the communication capability. Especially, the random matrix theory can be well mapped to the architecture of large array of antennas. The random matrix theory data model has ever been validated by \cite{zhang2014data} in a context of distributed sensing. In this paper, we extend this work to the massive MIMO testbed. In particular, we studied the function of multiple non-Hermitian random matrices and applied the variants to the experimental data collected on the massive MIMO testbed. The product of non-Hermitian random matrices shows encouraging potential in signal detection, that is motivated for spectrum sensing and network monitoring. We also present two concrete applications that are demonstrated on our testbed using the massive MIMO system as big data system. From the two applications, we foresee that, besides signal detection, the random-matrix based big data analysis will drive more mobile applications in the next generation wireless network.

\newtheorem{defi}{Definition}
\newtheorem{theorem}{Theorem}[section]
\newtheorem{lemma}[theorem]{Lemma}

\section{Modeling for Massive Data}
Large random matrices are used models for the massive data arising from the monitoring of the  massive MIMO system. We give some tutorial remarks, to facilitate the understanding of the experimental results.
\subsection{Data Modeling with Large Random Matrices}
Naturally, we assume $n$ observations of $p$-dimensional random vectors ${{\mathbf{x}}_1},...,{{\mathbf{x}}_n} \in {\mathbb{C}^{p \times 1}}.$ We form the data matrix ${\mathbf{X}} = \left( {{{\mathbf{x}}_1},...,{{\mathbf{x}}_n}} \right) \in {\mathbb{C}^{p \times n}}, $ which naturally, is a random matrix due to the presence of ubiquitous noise.  In our context, we are interested in the practical regime $p=100-1,000,$ while $n$ is assumed to be arbitrary.  The possibility of arbitrary sample size $n$ makes the classical statistical tools infeasible. We are asked to consider the asymptotic regime~\cite{qiu2012cognitive,Qiu_WicksBook2013,QiuAntonik2014Wiley,Qiu2016MassiveDataAnalysis} 
\begin{equation}
\label{eq(doubleasymp):def}%~\eqref{eq(doubleasymp):def}
	p \to \infty ,n \to \infty ,p/n \to c \in \left( {0,\infty } \right),
\end{equation}
while the classical regime~\cite{anderson2003introduction} considers 
\begin{equation}
\label{eq(doubleclassical):def}%~\eqref{eq(doubleclassical):def}
	p{\text{ fixed}},n \to \infty ,p/n \to 0.
\end{equation}
Our goal is to reduce massive data to a few statistical parameters. The first step often involves the covariance matrix estimation using the sample covariance estimator 
\begin{equation}
{\mathbf{S}} = \frac{1}{n}{\mathbf{X}}{{\mathbf{X}}^H} = \frac{1}{n}\sum\limits_{i = 1}^n {{{\mathbf{x}}_i}{\mathbf{x}}_i^H}  \in {\mathbb{C}^{p \times p}},
\end{equation}
that is a sum of rank-one random matrices~\cite{tropp2015introduction}. The sample covariance matrix estimator is the maximum likelihood estimator (so it is optimal) for the classical regime~\eqref{eq(doubleclassical):def}. However, for the asymptotic regime~\eqref{eq(doubleasymp):def}, this estimator is \textit{far from optimal}. We still use this estimator due to its special structure. See~\cite{qiu2012cognitive,Qiu_WicksBook2013,QiuAntonik2014Wiley,Qiu2016MassiveDataAnalysis} for modern alternatives to this fundamental algorithm. For brevity, we use the sample covariance estimator throughout this paper.

\subsection{Non-Hermitian Free Probability Theory}
Once data are modeled as large random matrices, it is natural for us to introduce the non-Hermitian random matrix theory into our problem at hand. Qiu's book~\cite{QiuAntonik2014Wiley} gives an exhaustive account of this subject in an independent chapter, from a mathematical view. This paper is complementary to our book~\cite{QiuAntonik2014Wiley} in that we bridge the gap between theory and experiments. We want to understand how accurate this theoretical model becomes for the real-life data. See Section~\ref{sect:RandomMatrixTheoreticalDataModel} for details. 

 Roughly speaking, large random  matrices can be treated as free matrix-valued random variables. ``Free'' random variables can be understood as independent random variables.  The matrix size must be so large that the asymptotic theoretical results are valid. It is of central interest to understand this  finite-size scaling in this paper.  

\section{Distributed Spectrum Sensing}
Now we are convinced that large random matrices are valid for experimental data modeling. The next natural question is to test whether the signal or the noise  is present in the data. Both networking monitoring and spectrum sensing can be formulated as a matrix hypothesis testing problem for anomaly detection.
\subsection{Related Work}
Specifically, consider the $n$ samples ${{\mathbf{y}}_1},...,{{\mathbf{y}}_n},$ drawn from a $p$-dimensional complex Gaussian distribution with covariance matrix ${\mathbf{\Sigma }}.$ We aim to test the hypothesis: \[{\mathcal{H}_0}:{\mathbf{\Sigma }} = {{\mathbf{I}}_p}.\]
This test has been studied extensively in classical settings (i.e., $p$ fixed, $n\to \infty$), first in
detail in~\cite{mauchly1940significance}. Denoting the sample covariance by ${{\mathbf{S}}_n} = \frac{1}{n}\sum\limits_{i = 1}^n {{{\mathbf{y}}_i}{\mathbf{y}}_i^H} ,$ the LRT is based
on the linear statistic (see Anderson (2003)~\cite[Chapter 10]{anderson2003introduction})
\begin{equation}
\label{eqdef:LRTlinearstatistic} %~\eqref{eqdef:LRTlinearstatistic}
	L = \operatorname{Tr} \left( {{{\mathbf{S}}_n}} \right) - \ln \left( {\det {{\mathbf{S}}_n}} \right) - p.
\end{equation}

Under ${\mathcal{H}_0},$ with $p$ fixed, as $n\to \infty,$ $nL$ is well known to follow a $\chi^2$ distribution. However, with high-dimensional data for which the dimension $p$ is large and comparable to the sample size $n,$ the $\chi^2$ approximation is no longer valid. A correction to the LRT is done in Bai, Jiang, Yao and Zheng (2009)~\cite{bai2009corrections} on  large-dimensional covariance matrix by random matrix theory. In this case, a better approach
is to use results based on the double-asymptotic given by Assumption 1. Such a study
has been done first under ${\mathcal{H}_0}$ and later under the spike alternative ${\mathcal{H}_1}.$ More specifically, under ${\mathcal{H}_0},$ this was presented in~\cite{bai2009corrections}  using a CLT framework established in Bai and Silverstein (2004)~\cite{Bai2004}. Under
``${\mathcal{H}_1}: {\mathbf{\Sigma }}$ has a spiked covariance structure as in Model A'', this problem was addressed only very recently in the independent works,~\cite{wang2014note} and~\cite{onatski2013asymptotic}. We point out that~\cite{wang2014note} (see also~\cite{onatski2014signal}) considered a generalized problem which allowed for multiple spiked eigenvalues. The result in~\cite{wang2014note} was again based
on the CLT framework of Bai and Silverstein (2004)~\cite{Bai2004}, with their derivation requiring the calculation of contour integrals. The same result was presented in~\cite{onatski2013asymptotic}, in this case making use of sophisticated tools of contiguity and Le Cam's first and third lemmas~\cite{vanderVaart1998AS}. 
\subsection{Spiked Central Wishart Matrix}

Our problem is formulated as
  \begin{equation}
	\label{eqdef:spikedcentralWishart }%~\eqref{eqdef:spikedcentralWishart }
	 \begin{gathered}
  {\mathcal{H}_0}:{\mathbf{\Sigma }} = {{\mathbf{I}}_p} \hfill \\
  {\mathcal{H}_1}:{\mathbf{\Sigma }} \in {\text{Model A: Spiked central Wishart}}{\text{.}} \hfill \\ 
\end{gathered}
 \end{equation}

\textit{Model A: Spiked central Wishart:} Matrices with distribution $\mathcal{C}{\mathcal{W}_p}\left( {n,{\mathbf{\Sigma }},{{\mathbf{0}}_{p \times p}}} \right)\left( {n \geqslant p} \right),$ where ${\mathbf{\Sigma }}$ has multiple distinct ``spike'' eigenvalues $1 + {\delta _1} >  \cdots  > 1 + {\delta _r},$ with ${\delta _r} > 0$ for all $1\le k \le r,$ and all other eigenvalues equal to 1.

\textbf{Assumption 1.} $n,p \to \infty {\text{ such that }} n/p\to c \geqslant 1.$

\begin{theorem}[Passemier, McKay and Chen (2014)~\cite{passemier2014hypergeometric}] 
\label{theorem1:passemier2014hypergeometric}
Consider Model A  and define 
\begin{equation}
\label{eq(3):passemier2014asymptotic} %~\eqref{eq(3):passemier2014asymptotic}
	a = {\left( {1 - \sqrt c } \right)^2},\;\;\;{\kern 1pt} b = {\left( {1 + \sqrt c } \right)^2}.
\end{equation}
Under Assumption 1, for an analytic function $f:\mathcal{U} \to \mathbb{C}$ where $\mathcal{U}$ is an open subset of the complex plane which contains $[a, b],$ we have\[\sum\limits_{i = 1}^p {f\left( {\frac{{{\lambda _i}}}{p}} \right)}  - p\mu \xrightarrow{\mathcal{L}}\mathcal{N}\left( {\sum\limits_{\ell  = 1}^r {\bar \mu \left( {{z_{0,\ell }}} \right)} ,{\sigma ^2}} \right),\]
where 
\begin{equation}
	\mu  = \frac{1}{{2\pi }}\int_a^b {f\left( x \right)\frac{{\sqrt {\left( {b - x} \right)\left( {x - a} \right)} }}{x}} dx
\end{equation}
\begin{equation}
	{\sigma ^2} = \frac{1}{{2{\pi ^2}}}\int_a^b {\frac{{f\left( x \right)}}{{\sqrt {\left( {b - x} \right)\left( {x - a} \right)} }}\left[ {\mathcal{P}\int_a^b {\frac{{f'\left( y \right)\sqrt {\left( {b - y} \right)\left( {y - a} \right)} }}{{x - y}}} dy} \right]} dx
\end{equation}
with these terms independent of the spikes. The spike-dependent terms $\bar \mu \left( {{z_{0,\ell }}} \right),1 \leqslant \ell  \leqslant r$
admit
\begin{equation}
	\bar \mu \left( {{z_{0,\ell }}} \right) = \frac{1}{{2\pi }}\int_a^b {\frac{{f\left( x \right)}}{{\sqrt {\left( {b - x} \right)\left( {x - a} \right)} }}\left[ {\frac{{\sqrt {\left( {{z_{0,\ell }} - a} \right)\left( {{z_{0,\ell }} - b} \right)} }}{{{z_{0,\ell }} - x}} - 1} \right]} dx
\end{equation}
where \[{z_{0,\ell }} = \left\{ {\begin{array}{*{20}{c}}
   {\frac{{\left( {1 + c{\delta _\ell }} \right)\left( {1 + {\delta _\ell }} \right)}}{{{\delta _\ell }}},} & {{\text{ for Model A}}}  \\ 
   {\frac{{\left( {1 + {\nu _\ell }} \right)\left( {1 + {\nu _\ell }} \right)}}{{{\nu _\ell }}},} & {{\text{for Model B}}}  \\ 
\end{array} } \right..\]
The branch of the square root ${\sqrt {\left( {{z_{0,\ell }} - a} \right)\left( {{z_{0,\ell }} - b} \right)} }$ is chosen.
\end{theorem}
As an application of Theorem~\ref{theorem1:passemier2014hypergeometric} for Model A, we consider the classical LRT that the
population covariance matrix is the identity, under a rank-one spiked population alternative.

Here, we will adopt our general framework to recover the same result as~\cite{wang2014note} and~\cite{onatski2013asymptotic} very efficiently, simply by calculating a few integrals. Under ${\mathcal{H}_1},$ as before we denote by
$1 +\delta$ the spiked eigenvalue of ${\mathbf{\Sigma }}.$ Since $n{{\mathbf{S}}_n} \sim \mathcal{C}{\mathcal{W}_p}\left( {n,{\mathbf{\Sigma }},{{\mathbf{0}}_{p \times p}}} \right),$ we now apply Theorem~\ref{theorem1:passemier2014hypergeometric} for the case of Model A to the function\[{f_L}\left( x \right) = \frac{x}{c} - \ln \left( {\frac{x}{c}} \right) - 1.\]
Let ${\lambda _i},1 \leqslant i \leqslant p,$ be the eigenvalues of $n{{\mathbf{S}}_n}.$ Since the domain of definition of $f_L$ is $(0,\infty),$ we assume that $c > 1$ to ensure $a > 0$ (see~\eqref{eq(3):passemier2014asymptotic}). Then, under Assumption 1, \[L = \sum\limits_{i = 1}^p {{f_L}\left( {\frac{{{\lambda _i}}}{p}} \right)\mathop  \to \limits^\mathcal{L} } \mathcal{N}\left( {p\mu  + \bar \mu ,{\sigma ^2}} \right),\] \[L = \sum\limits_{i = 1}^p {{f_L}\left( {\frac{{{\lambda _i}}}{p}} \right)\mathop  \to \limits^\mathcal{L} } \mathcal{N}\left( {p\mu  + \bar \mu \left( {{z_{0,1}}} \right),{\sigma ^2}} \right),\]
where $r=1$ is used for one spike to obtain \[\mu  = 1 + \left( {c - 1} \right)\ln \left( {1 - {c^{ - 1}}} \right),\;\;\;{\kern 1pt} {\sigma ^2} =  - {c^{ - 1}}\ln \left( {1 - {c^{ - 1}}} \right)\] with the spike-dependent term \[\bar \mu  = {\delta _1} - \ln \left( {1 + {\delta _1}} \right).\] The special case of one spike is also considered in~\cite{passemier2014asymptotic}. These results are in agreement with~\cite{wang2014note} and~\cite{onatski2013asymptotic}.

\subsection{Distributed Streaming Data}
 For each server,~equation~\eqref{eqdef:spikedcentralWishart } formulates the testing problem. How do we formulate this problem when the  data are spatially  distributed across $N$ servers? Our proposed algorithm is as follows:               \textbf{Algorithm 1}
 \begin{enumerate}
	 \item The $i$-th server  computes the sample covariance matrix ${\bf S}_i,i=1,...,N.$
	\item  The $i$-th server computes the linear statistic \[L_i = \operatorname{Tr} \left( {{{\mathbf{S}}_i}} \right) - \ln \left( {\det {{\mathbf{S}}_i}} \right) - p, i=1,...,N.\]
	\item The $i$-th server communicates the linear statistic $L_i,i=1,...,N$ to one server that acts as the coordinator.
	\item Finally, the coordinator server obtains the linear statistic $L_i,i=1,...,N$ via communication and sum up the values $L _{D}= {L_1} +  \cdots  + {L_N}.$
	\item All the above computing and communication are done in in parallel.
 \end{enumerate}

The communication burden is very low. The central ingredient of Algorithm 1 is to exploit the Central Limit Theorem of the used linear statistic $L$ defined in~\eqref{eqdef:LRTlinearstatistic}.  By means of Theorem~\ref{theorem1:passemier2014hypergeometric}, we have \[L = \sum\limits_{i = 1}^p {f\left( {\frac{{{\lambda _i}}}{p}} \right)} \xrightarrow{\mathcal{L}}\mathcal{N}\left( {p\mu  + \sum\limits_{\ell  = 1}^r {\bar \mu \left( {{z_{0,\ell }}} \right)} ,{\sigma ^2}} \right).\] Since $L_1,...,L_N$ are Gaussian random variables, the sum of Gaussian random variables are also Gaussian; thus ${L_D} = {L_1} +  \cdots  + {L_N}$ is also Gaussian, denoted as $\mathcal{N}\left( {{\mu _D},\sigma _D^2} \right).$

The false alarm probability for the linear statistic can be obtained using standard procedures. If $L_D>\gamma,$ the signal is present; otherwise, the signal does not exist. The false alarm probability is \[\begin{array}{*{20}{c}}
   {{P_{fa}} = \mathbb{P}\left( {L > \gamma \left| {{\mathcal{H}_0}} \right.} \right)} \hfill & { = \mathbb{P}\left( {\frac{{L - {\mu _D}}}{{{\sigma _D}}} > \frac{{\gamma  - {\mu _D}}}{{{\sigma _D}}}\left| {{\mathcal{H}_0}} \right.} \right)} \hfill  \\ 
   {} \hfill & { = \int\limits_{\frac{{L - {\mu _D}}}{{{\sigma _D}}}}^\infty  {\frac{1}{{\sqrt {2\pi } }}\exp \left( { - {t^2}/2} \right)dt} } \hfill  \\ 
   {} \hfill & { = Q\left( {\frac{{L - {\mu _D}}}{{{\sigma _D}}}} \right)} \hfill  \\ 
\end{array} \]where $Q\left( x \right) = \int_x^\infty  {\frac{1}{{\sqrt {2\pi } }}\exp \left( { - {t^2}/2} \right)dt.} $
For a desired false-alarm rate $\varepsilon,$ the associated threshold should be chosen such that
\[\gamma  = {\mu _D} + \frac{1}{{{\sigma _D}}}{Q^{ - 1}}\left( \varepsilon  \right).\]
To predict the detection probability, we need to know the distribution of $\xi$ under ${\mathcal H}_1,$ which has been obtained using Theorem~\ref{theorem1:passemier2014hypergeometric}. The detection probability is calculated as\[\begin{array}{*{20}{c}}
   {{P_d} = \mathbb{P}\left( {{L_D} > \gamma \left| {{\mathcal{H}_1}} \right.} \right)} \hfill & { = \mathbb{P}\left( {\frac{{\xi  - {\mu _D}}}{{{\sigma _D}}} > \frac{{\gamma  - {\mu _D}}}{{{\sigma _D}}}\left| {{\mathcal{H}_1}} \right.} \right)} \hfill  \\ 
   {} \hfill & { = Q\left( {\frac{{{L_D} - {\mu _D}}}{{{\sigma _D}}}} \right).} \hfill  \\ 
\end{array} \]

\section{Massive MIMO Testbed and Data Acquisition}

\subsection{System Architecture and Signal Model}
\begin{figure}[h]
 \centering
\includegraphics[width=3.4in]{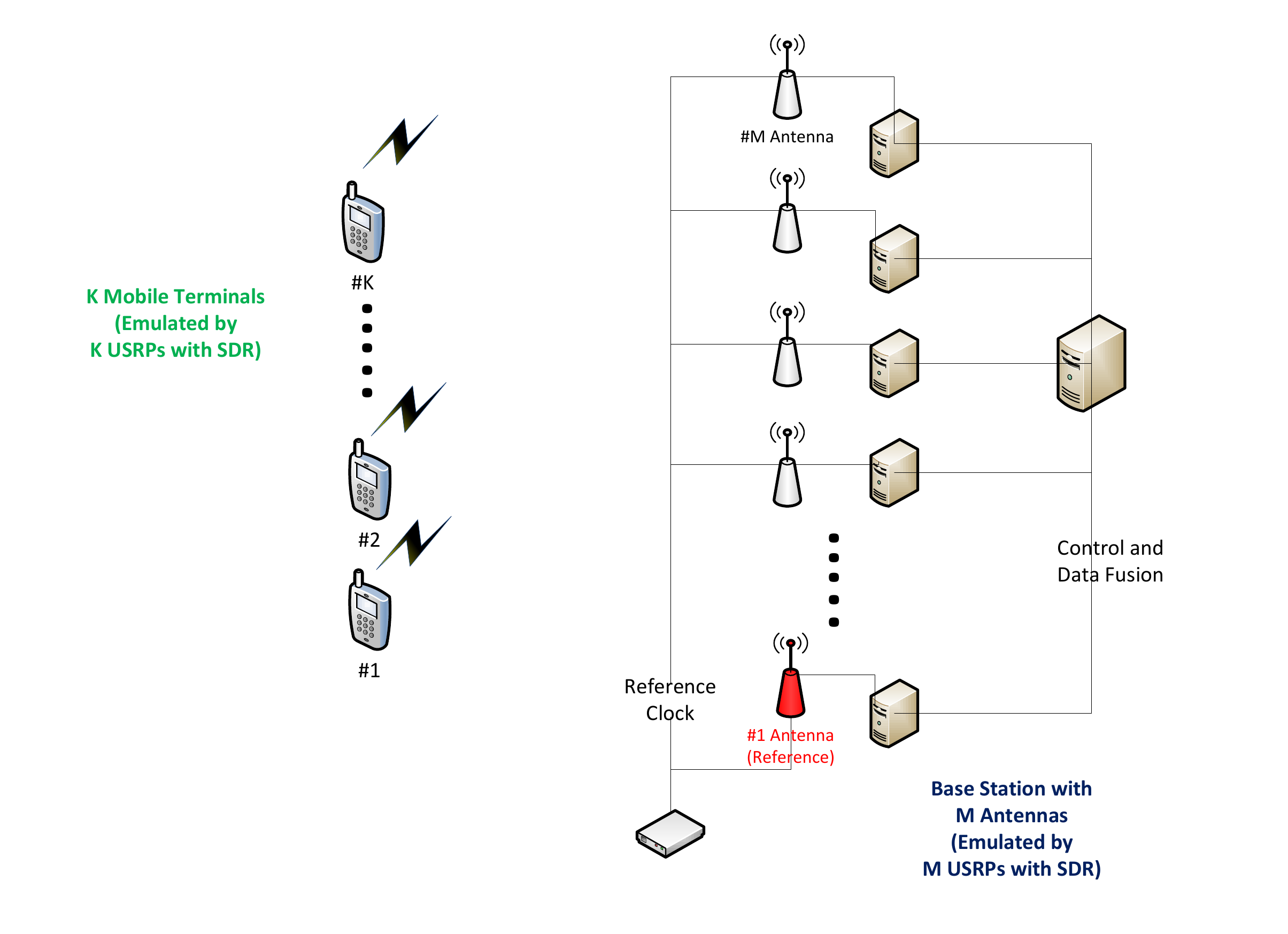}
\caption{System Architecture of Multi-User Massive MIMO Testbed.}
\label{sysarch}
\end{figure}

%We setup a software defined radio (SDR) based wireless sensor network with up to 70 Nodes. 
%The data for noise only or commercial signal at 869.5MHz are collected respectively.
The system architecture of the testbed is as 
Fig.~\ref{sysarch}.

The general software-defined radio (SDR) universal software radio peripheral (USRP) platform is used to emulate the base station antenna in our testbed. We deployed up to 70 USRPs and 30 high performance PCs to work collaboratively as an large antenna array of the massive MIMO base station. These USRPs are well clock synchronized by an AD9523 clock distribution board. The system design of this testbed can be found in \cite{zhang2014massive}.

Our testbed has demonstrated initial capabilities as below:

\paragraph{Channel Reciprocity for Channel Measurement}

Channel matrix measurement is a critical task for Multi-User Massive MIMO system. 
For the antenna $i$ and $j$, if the uplink and downlink work in TDD mode, the channel reciprocity will be useful for the pre-coding in MIMO system.
Channel reciprocity means $h_{i,j} = h_{j,i}$ if $h_{i,j}$ represents the air channel from antenna $i$ to antenna $j$ and vice versa. 

\begin{figure}
\centering
\includegraphics[width=3.4in]{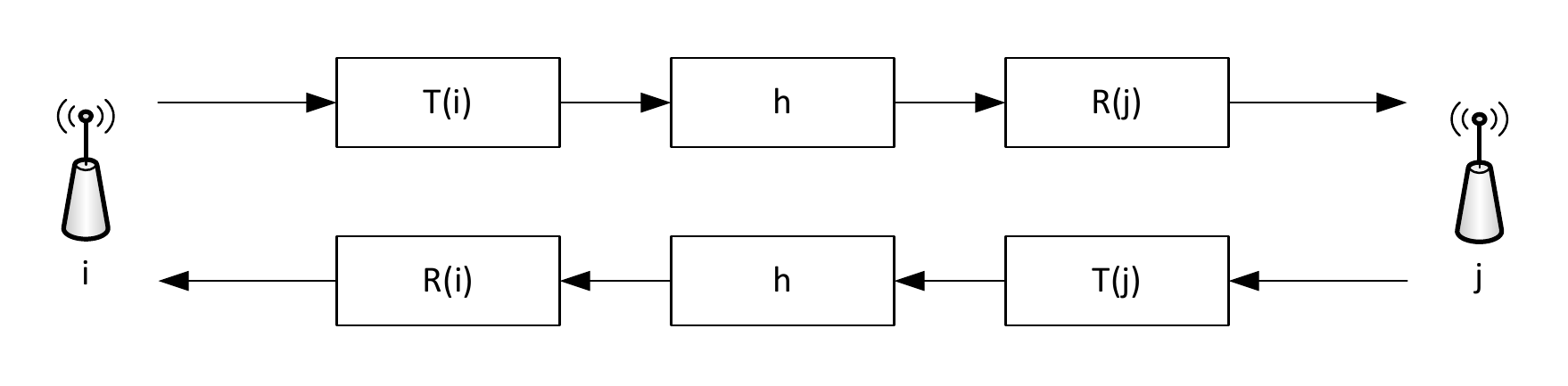}
\caption{Reciprocity mode for TDD channel}
\label{reciprocity}
\end{figure}

Given the $h$ is the air channel between antenna $i$ and $j$, the measured channel $h_{i,j}$ and $h_{j,i}$ follow the model depicted as Fig.~\ref{reciprocity}, where
where $T\left( i \right)$, ${R\left( j \right)}$, $R\left( i \right)$, $T\left( j \right)$ represent the effect from circuits like upper/down conversion, filters, etc., for
both the upper and down links.

Thus we have
\begin{equation}
\begin{gathered}
  {h_{i,j}} = T\left( i \right) \cdot h \cdot R\left( j \right) \hfill \\
  {h_{i,j}} = R\left( i \right) \cdot h \cdot T\left( j \right) \hfill \\ 
\end{gathered} 
\end{equation}

Usually, the relative calibration is sufficient for the pre-coding as we have
\begin{equation}
\frac{{{h_{i,j}}}}{{{h_{j,i}}}} = \frac{{T\left( i \right) \cdot R\left( j \right)}}{{R\left( i \right) \cdot T\left( j \right)}}
\end{equation}
which is constant in ideal situation.

Channel reciprocity described above includes the circuits impact. Our measurement shows that ratio ${{{h_{i,j}}} \mathord{\left/
 {\vphantom {{{h_{ij}}} {{h_{ji}}}}} \right.
 \kern-\nulldelimiterspace} {{h_{ji}}}}$  between the
downlink and uplink channel frequency response for antenna $i$ and $j$ is almost constant. For example, we collect 3 rounds of data within a time duration that the channel can be regarded as static. Thus
3 such ratios are obtained for a specified link between USPR node transmitting antenna $3$ and receiving antenna $22$. The absolute value of 3 ratios are 
1.2486, 1.22, 1.2351 respectively. 

\paragraph{Massive Data Acquisition for Mobile Users or Commercial Networks}

Consider the time evolving model described as below:
Let $N$ be the number of antennas at base station. All the antennas start sensing at the same time.
Every time, on each antenna, a time series of with samples length $T$ is captured and 
denoted as ${x_i} \in {\mathbb{C}^{1 \times T}},i = 1, \ldots ,N$. Then a random matrix
from $N$ such vectors are formed as:
\begin{equation}
{{\rm X}_j} = {\left[ {\begin{array}{*{20}{c}}
  {{x_1}} \\ 
  {{x_2}} \\ 
   \vdots  \\ 
  {{x_N}} 
\end{array}} \right]_{N \times T}}
\end{equation}
where, $j = 1, \cdots ,L$. Here $L$ means we repeat the sensing procedure with $L$ times. Then $L$ such
random matrices are obtained. In the following sections, we are interested in variant random matrix theoretical data models, including the product of the $L$ random matrices and their geometric/arithmetic mean. We call it time evolving approach.

Besides the time evolving approach, we can also use a different data format to form random matrix. Suppose we select $n$ receivers at Massive MIMO base station. At each receiver, we collect $N \times T$ samples to get a random matrix ${{\rm X}_i} \in {\mathbb{C}^{N \times T}}$ with $i = 1, \cdots ,n$.
Similarly, we are interested in the functions of these random matrices. We call it space distributed approach.

In the next section, we specify which approach is used to form the random matrix when a certain theoretical model is used. 

\section{Random Matrix Theoretical Data Model and Experimental Validation}
\label{sect:RandomMatrixTheoreticalDataModel}
We are interested in the eigenvalue distribution for every data model. The results obtained from the experimental data are compared with theoretical distribution (if exists). The experimental data come from noise-only case and signal-present case. Our testbed captures the commercial signal data at $869.5MHz$.
\subsection{Product of non-Hermitian random matrix}
\label{proNH}
The eigenvalue distribution for the product of non-Hermitian random matrix, so far, gives us the best visible information
to differentiate the situations of noise only and signal present. Here the timing evolving approach is used.
Denote the product of non-Hermitian random matrix as:
\begin{equation}
{\text{Z}} = \prod\limits_{j = 0}^L {{{\rm X}_j}} 
\end{equation}
In the experiment, $L$ is adjustable. In addition, a number of such $Z$ are captured with time evolving, to investigate
if the pattern is changing or not with time. Every $Z$ could be regarded as one snapshot. For both the noise and signal 
experiment, we took $10$ snapshots. All the 10 snapshots are put together to show eigenvalue distribution more clearly.

\subsubsection{Eigenvalue Distributions for Noise-Only and Signal-Present}

Firstly, we visualize the eigenvalue distribution on the complex plane to see the difference for the cases of noise-only and signal-present.

\begin{figure}
 \centering
\includegraphics[width=3.4in]{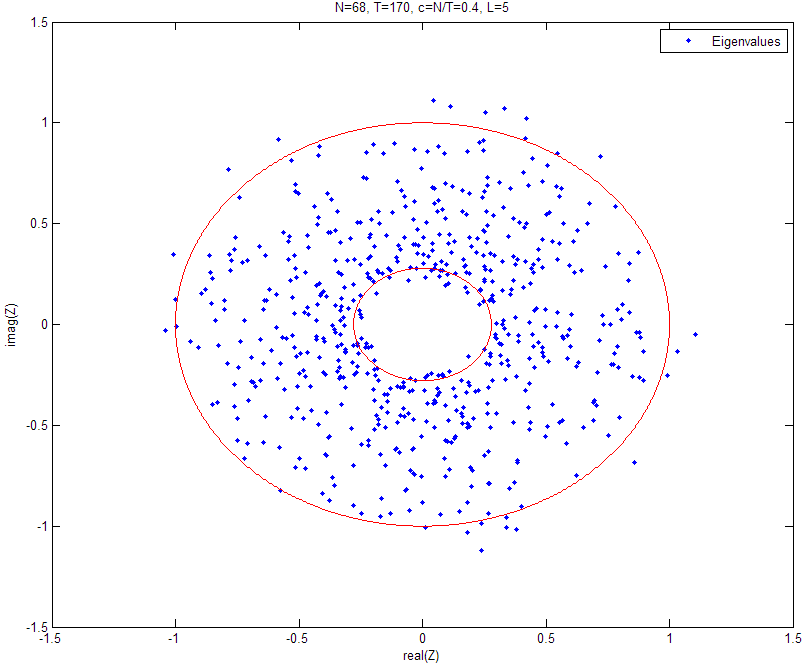}
\caption{The eigenvalue distribution for product of non-Hermitian random matrix, noise only, all snapshots.}
\label{noise_product_allshot}
\end{figure}

\textbf{Noise Only:}
If the eigenvalue distribution for all the snapshots are put together, we see Fig.~\ref{noise_product_allshot}, in which the red circle represents the ``Ring Law''.

\begin{figure}
 \centering
\includegraphics[height=2.5in]{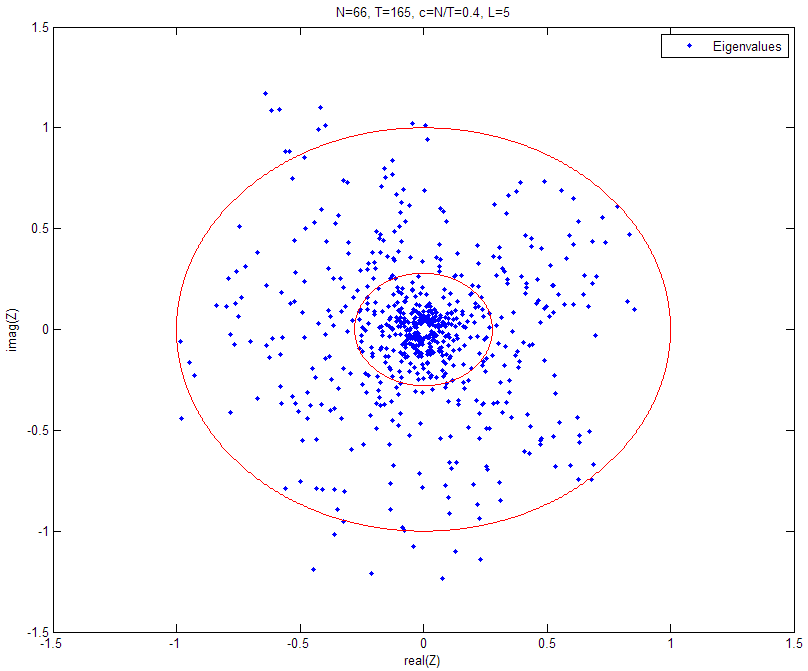}
\caption{The eigenvalue distribution for product of non-Hermitian random matrix, signal present, all snapshots.}
\label{signal_product_allshot}
\end{figure}

\textbf{Signal Present:}
If putting together the eigenvalues of all snapshots, we see Fig.~\ref{signal_product_allshot}, in which the inner radius of the eigenvaule distribution is smaller than that of the ring law.

We also use the probability density diagram to show the difference between noise only and signal
present cases, with different $L$. The theorem \ref{ring} actually gives the theoretical values of the inner radius and outer radius of the ring law.
\begin{theorem}
\label{ring}
The empirical eigenvalue distribution of  $N \times T$ matrix ${\prod\limits_{i = 1}^L {{{\rm X}_i}} }$ converge
almost surely to the same limit given by
%\emph{(Lagrange's Theorem)}
\[{f_{\prod\limits_{i = 1}^L {{{\rm X}_i}} }}\left( \lambda  \right) = \left\{ {\;\begin{array}{*{20}{c}}
  {\frac{2}{{\pi cL}}{{\left| \lambda  \right|}^{{2 \mathord{\left/
 {\vphantom {2 L}} \right.
 \kern-\nulldelimiterspace} L} - 2}}\quad } \\ 
  0 
\end{array}} \right.\;\begin{array}{*{20}{c}}
  {{{\left( {1 - c} \right)}^{{L \mathord{\left/
 {\vphantom {L 2}} \right.
 \kern-\nulldelimiterspace} 2}}} \leqslant r \leqslant 1} \\ 
  {{\text{elsewhere}}} 
\end{array}\]
as $N$, $n$ $ \to \infty $ with the ratio $c = {N \mathord{\left/
 {\vphantom {N n}} \right.
 \kern-\nulldelimiterspace} n} \leqslant 1$ fixed.
\label{esd_product}

\end{theorem}

We are interested in the probability density of ${\left| \lambda  \right|}$. Let $r = \left| \lambda  \right|$, which is described in
Eq.~\ref{product_radius_pdf}, derived from the Theorem~\ref{esd_product}.
\begin{equation}
 {f_{\prod\limits_{i = 1}^L {{{\rm X}_i}} }}\left( r \right) = \left\{ {\begin{array}{*{20}{c}}
  {\frac{2}{{cL}}{r^{\frac{2}{L} - 1}}} \\ 
  0 
\end{array}} \right.\quad \begin{array}{*{20}{c}}
  {{{\left( {1 - c} \right)}^{{L \mathord{\left/
 {\vphantom {L 2}} \right.
 \kern-\nulldelimiterspace} 2}}} \leqslant r \leqslant 1} \\ 
  {elsewhere} 
\end{array}\;
\label{product_radius_pdf}
\end{equation}

\begin{figure}
 \centering
\includegraphics[width=3.4in]{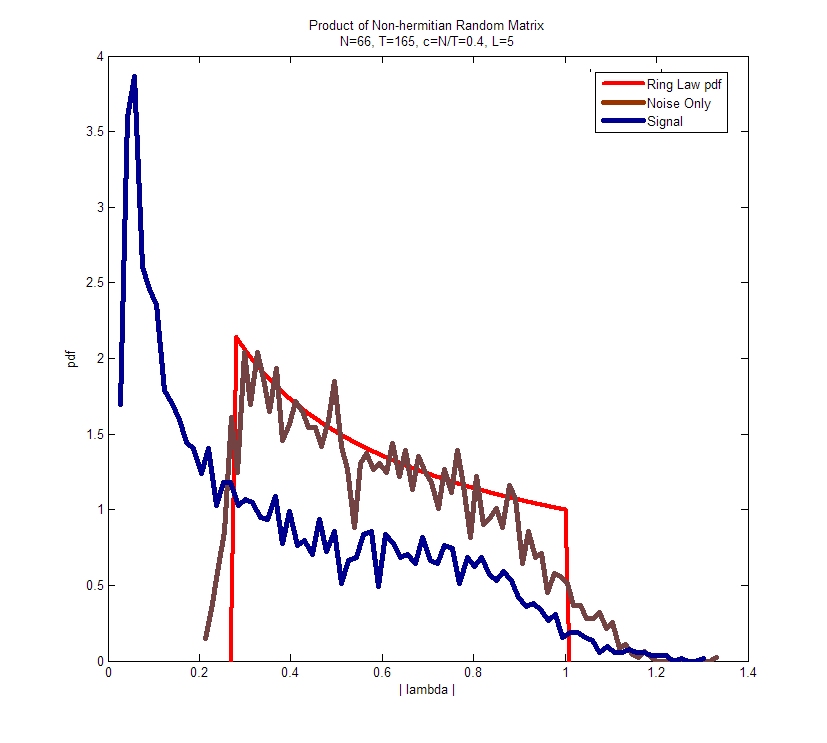}
\caption{Probability of eigenvalue for product of the non-Hermitian random matrix, both cases, with $L=5$.}
\label{pdf_ns_product_L5}
\end{figure}

\begin{figure}
 \centering
\includegraphics[width=3.4in]{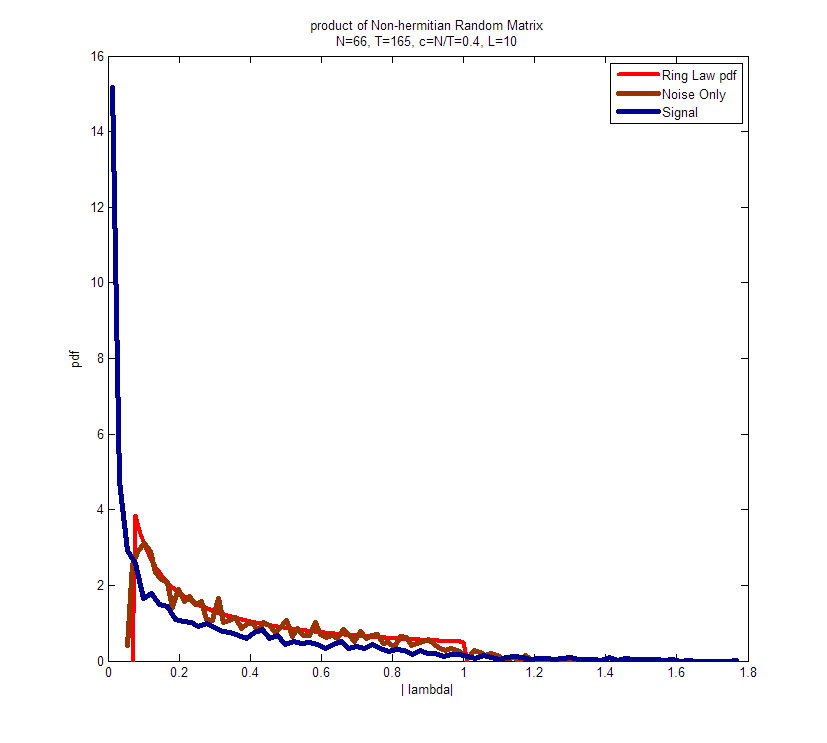}
\caption{Probability of eigenvalue for product of the non-Hermitian random matrix, both cases, with $L=10$.}
\label{pdf_ns_product_L10}
\end{figure}

The PDF is also shown in Fig.~\ref{pdf_ns_product_L5} and Fig.~\ref{pdf_ns_product_L10} with different $L$.

The above results show that eigenvalue distribution follows the ring law in this model for noise only case. The signal present case also has the ring law while the inner radius is much smaller than the noise only case, especially when $L$ is large. 

\subsubsection{Empirical Effect of $L$ to Differentiate Cases of Noise only and Signal Present}:

Regarding the product of non-Hermitian random matrices, the main difference observed in cases of noise only and signal present, is about the inner circle radius of the eigenvalue distribution. 

According to the ring law, the inner circle radius of the eigenvalue distribution for the noise only case, is constrained by Eq.~\ref{ringlaw}, which is a fixed value for a determined $L$ and $c$.
\begin{equation}
{r_{\text{inner}}} = {\left( {1 - c} \right)^{\frac{L}{2}}}
\label{ringlaw}
\end{equation}
Meanwhile, the radius shrinks for the case of the signal being present. In addition, for both cases, the inner circle radius decreases with increasing $L$. The question is whether it is easier to differentiate the two cases
when increasing the value $L$? 

For the same $L$, we define ${\left. {{M_{\text{noise}}}\left( L \right)} \right|_{r < {r_{_{\text{inner}}}}}}$ as the number of eigenvalues falling within the ring law inner circle, measured for the noise only case, and the ${\left. {{M_{\text{noise}}}\left( L \right)} \right|_{r < {r_{_{\text{inner}}}}}}$ as the number of eigenvalues falling within the inner circle of the ring law, measured for signal present case. Thus, we have a ratio
denoted as 
\begin{equation}
\rho \left( L \right) = \frac{{{{\left. {{M_{\text{noise}}}\left( L \right)} \right|}_{r < {r_{_{\text{inner}}}}}}}}{{{{\left. {{M_{\text{signal}}}\left( L \right)} \right|}_{r < {r_{_{\text{inner}}}}}}}}
\end{equation}
to represent the impact of $L$.

\begin{figure}
 \centering
\includegraphics[width=3.4in]{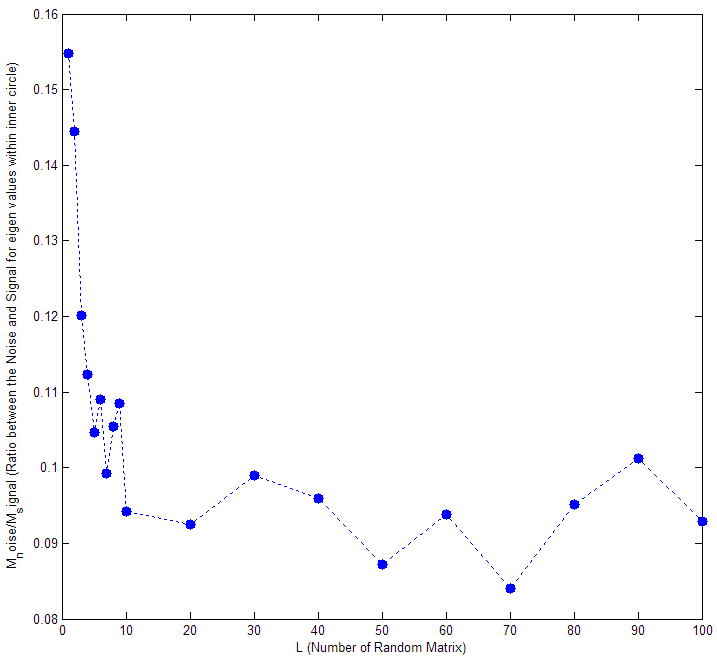}
\caption{Shrinking eigenvalue ratio within the ring law inner circle between the noise only and signal present
cases.}
\label{ratio_noise_signal}
\end{figure}

Fig.~\ref{ratio_noise_signal} show the trend of the ratio with increasing $L$. Generally, the ratio decreases with the increasing $L$, indicating that the larger $L$ brings better distance between the cases of noise only and 
signal present. However, the trend is very similar with the negative exponential function of $L$. When $L$ is greater than $10$, the ratio does not change much. 

\subsection{Geometric Mean}
Using the same data as last paragraph, the geometric mean of the non-Hermitian random matrix can be obtained as:
\begin{equation}
{\text{Z}} = {\left( {\prod\limits_{j = 0}^L {{{\rm X}_j}} } \right)^{{1 \mathord{\left/
 {\vphantom {1 L}} \right.
 \kern-\nulldelimiterspace} L}}}
\end{equation}

Time evolving approach is used here. In this experiment, we adjust the $L$ and the convergence is observed when $L$ is increased. All the diagrams below include 10 snapshots of data results. Basically, in this case, the eigenvalues converge to the outer unit
circle and are not changing much with increasing $L$.

\begin{figure}
 \centering
\includegraphics[width=3.4in]{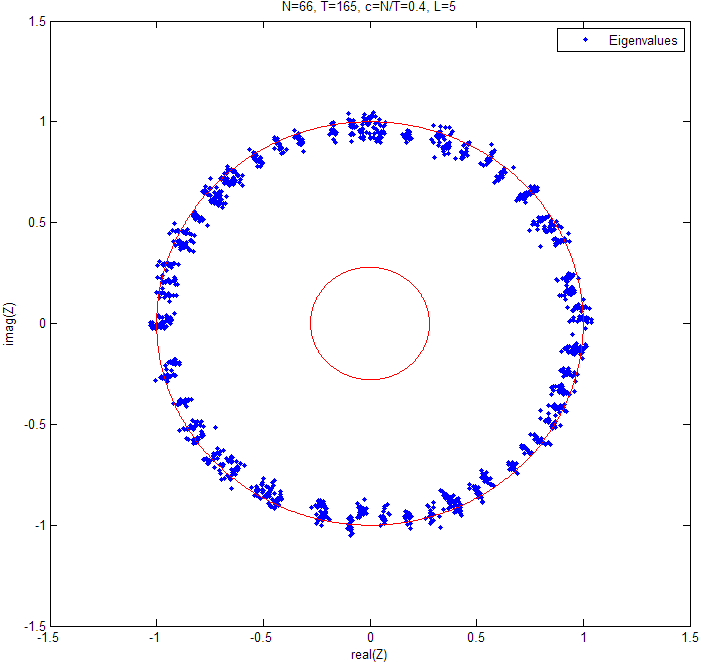}
\caption{The eigenvalue distribution for geometric mean of non-Hermitian random matrix, noise only, all snapshots, $L$=5.}
\label{noise_gmean_allshot_l5}
\end{figure}

\begin{figure}
 \centering
\includegraphics[width=3.4in]{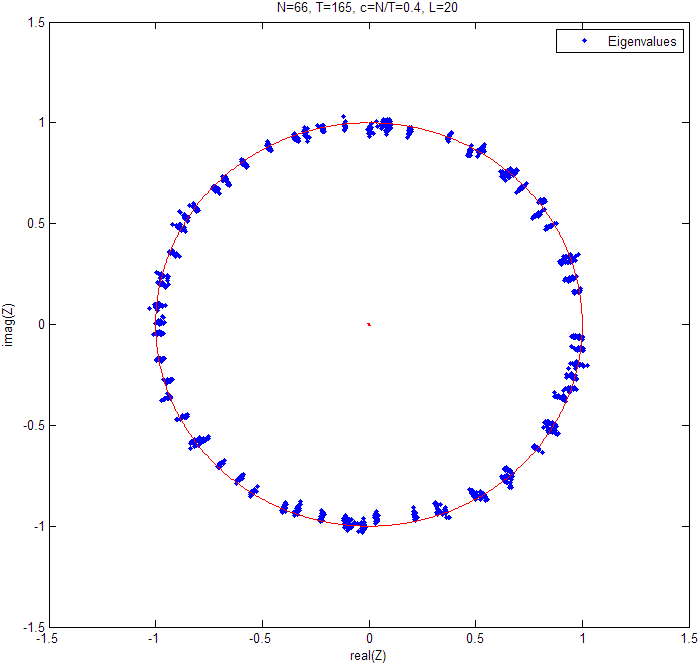}
\caption{The eigenvalue distribution for geometric mean of non-Hermitian random matrix, noise only, all snapshots, $L$=20.}
\label{noise_gmean_allshot_l20}
\end{figure}

\begin{figure}
 \centering
\includegraphics[width=3.4in]{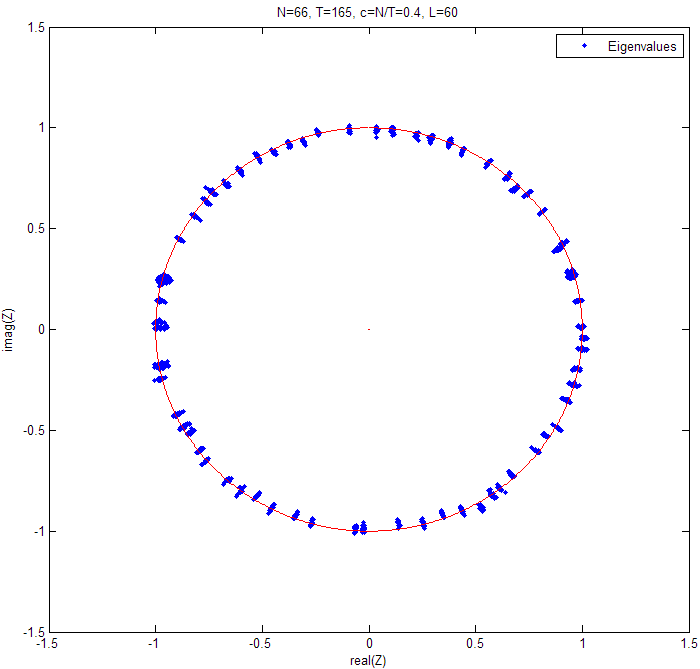}
\caption{The eigenvalue distribution for geometric mean of non-Hermitian random matrix, noise only, all snapshots, $L$=60.}
\label{noise_gmean_allshot_l60}
\end{figure}

\textbf{Noise Only}: 
Fig.~\ref{noise_gmean_allshot_l5} to Fig.~\ref{noise_gmean_allshot_l60} show the eigenvalue distribution of the
geometric mean for noise only situation. 

\begin{figure}
 \centering
\includegraphics[width=3.4in]{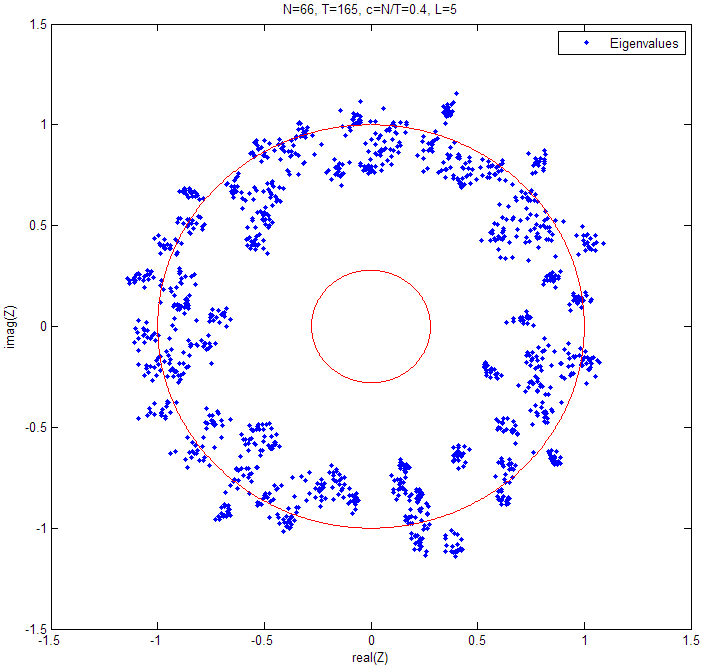}
\caption{The eigenvalue distribution for geometric mean of non-Hermitian random matrix, signal present, all snapshots, $L$=5.}
\label{signal_gmean_allshot_l5}
\end{figure}

\begin{figure}
 \centering
\includegraphics[width=3.4in]{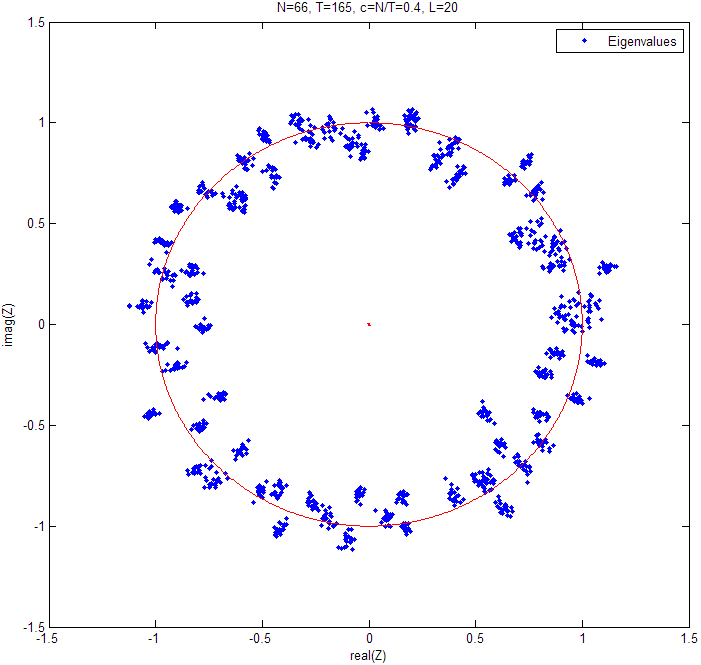}
\caption{The eigenvalue distribution for geometric mean of non-Hermitian random matrix, signal present, all snapshots, $L$=20.}
\label{signal_gmean_allshot_l20}
\end{figure}

\begin{figure}
 \centering
\includegraphics[width=3.4in]{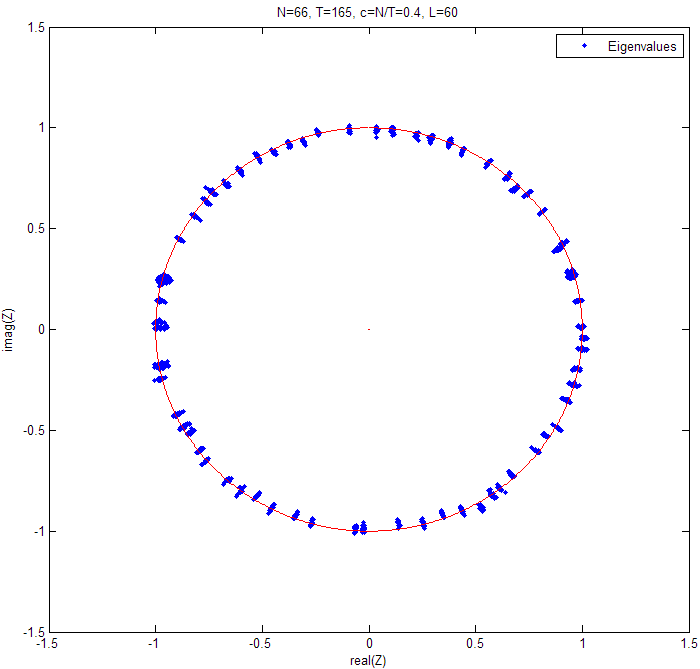}
\caption{The eigenvalue distribution for geometric mean of non-Hermitian random matrix, signal present, all snapshots, $L$=60.}
\label{signal_gmean_allshot_l60}
\end{figure}

\textbf{Signal Present}:
Fig.~\ref{signal_gmean_allshot_l5} to Fig.~\ref{signal_gmean_allshot_l60} show the eigenvalue distribution of the
geometric mean for signal situation. Different with noise case, the convergence of the eigenvalue is sensitive to the value of $L$. With bigger $L$, the distribution converges more to the unit circle.

\begin{figure}
 \centering
\includegraphics[width=3.4in]{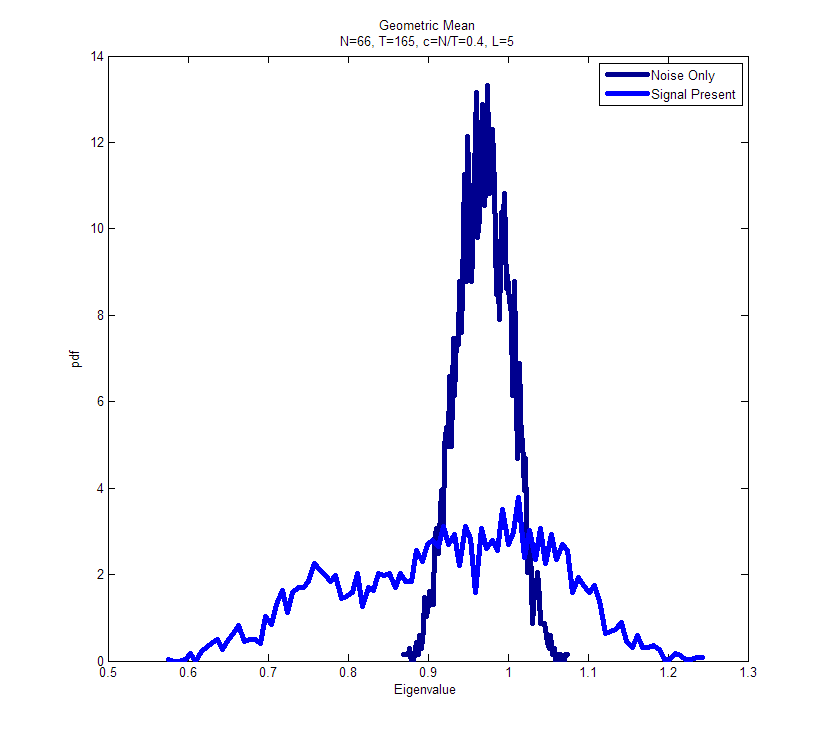}
\caption{Probability of eigenvalue for geometric mean of the non-Hermitian random matrix, both cases, with $L=5$.}
\label{pdf_ns_gmean_L5}
\end{figure}

\begin{figure}
 \centering
\includegraphics[width=3.4in]{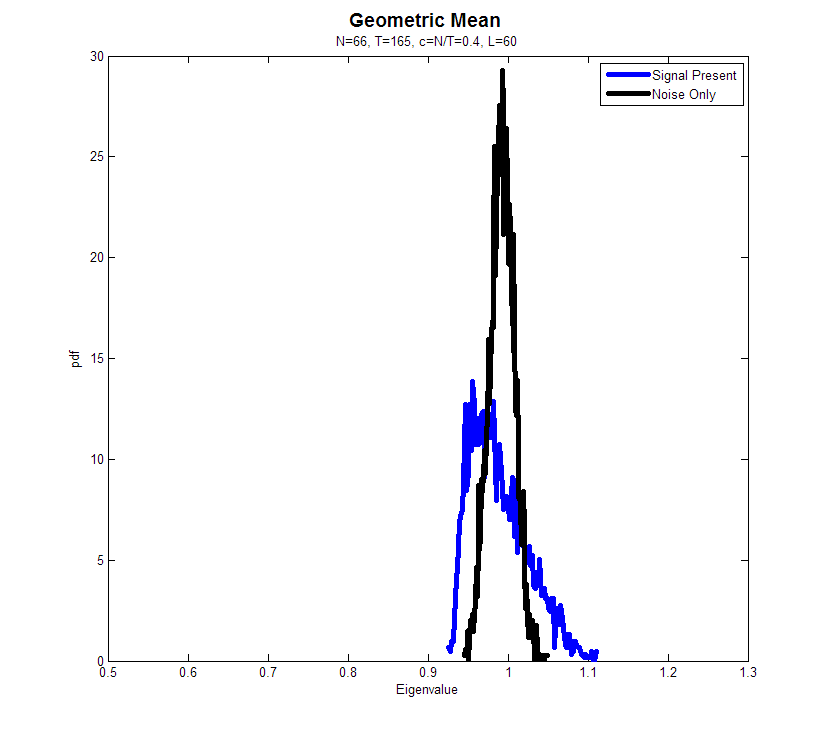}
\caption{Probability of eigenvalue for geometric mean of the non-Hermitian random matrix, both cases, with $L=60$.}
\label{pdf_ns_gmean_L60}
\end{figure}

We also show the PDF of the eigenvalue absolute values for geometric mean, in Fig.~\ref{pdf_ns_gmean_L5} and Fig.~\ref{pdf_ns_gmean_L60} with 
different $L$.

From all the visualized results for the Geometric mean model, we see that
\begin{itemize}
 \item the eigenvalue distribution is similar to the ring law, but the radius is not the same as product of non-Hermitian random matrices.
 \item the difference between inner radius and the outer radius, for the signal-present case, is larger than that for noise-only case.
 \item with $L$ increased, the ``ring'' is converged more to the outer circle. The absolute difference between noise-only and signal-present is actually not get larger with increasing $L$.
\end{itemize}

\subsection{Arithmetic Mean:}  
The arithmetic mean of the non-Hermitian random matrix is defined as 
\begin{equation}
{\text{Z = }}\frac{1}{L}\left( {\sum\limits_j^L {{{\rm X}_j}} } \right)
\end{equation}
For both the noise-only and signal-present cases, we adjust the value
of $L$ to see the effect. We select $L=5,20,100$.

\begin{figure}
 \centering
\includegraphics[width=3.4in]{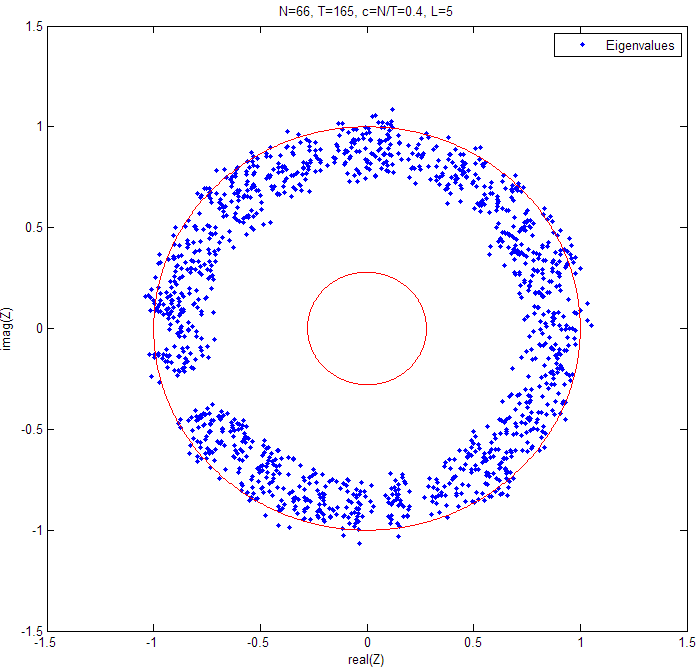}
\caption{The eigenvalue distribution for arithmetic mean of non-Hermitian random matrix, noise only, $L$=5.}
\label{noise_amean_l5}
\end{figure}

\begin{figure}
 \centering
\includegraphics[width=3.4in]{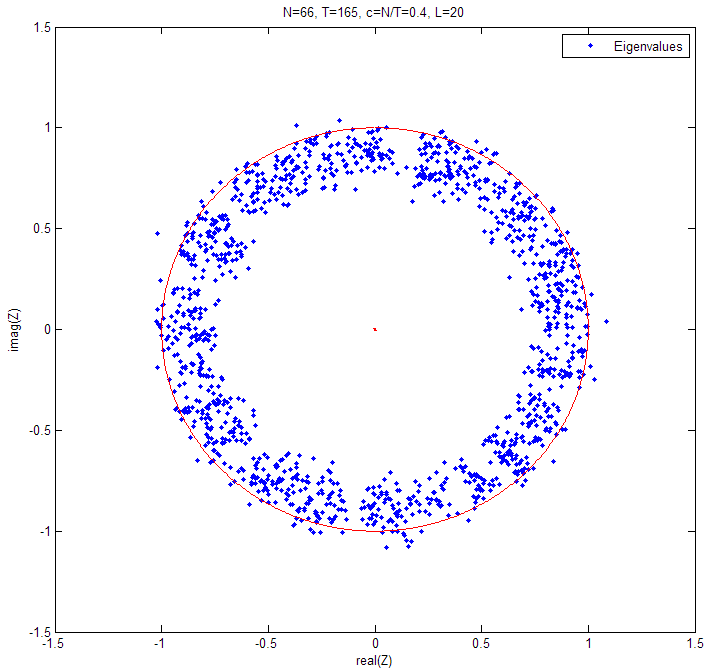}
\caption{The eigenvalue distribution for arithmetic mean of non-Hermitian random matrix, noise only, $L$=20.}
\label{noise_amean_l20}
\end{figure}

\begin{figure}
 \centering
\includegraphics[width=3.4in]{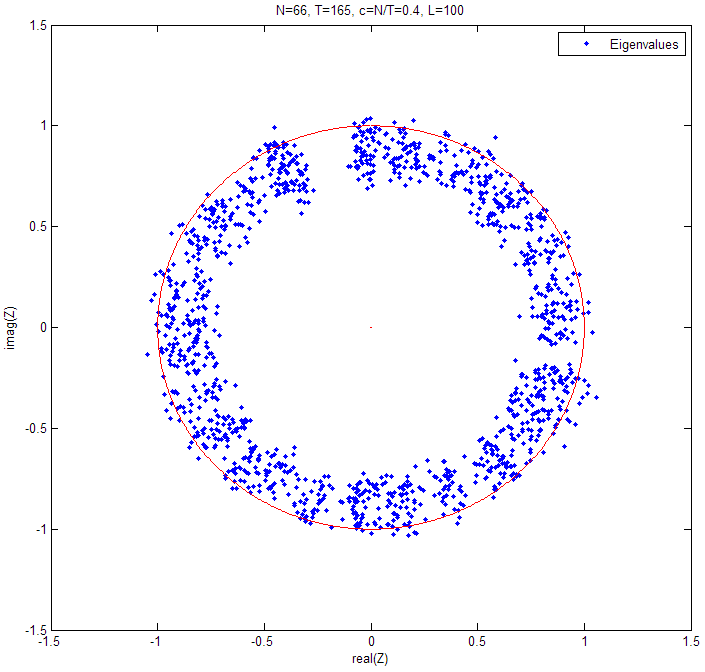}
\caption{The eigenvalue distribution for arithmetic mean of non-Hermitian random matrix, noise only, $L$=100.}
\label{noise_amean_l100}
\end{figure}

\textbf{Noise Only}: Fig.~\ref{noise_amean_l5} to Fig.~\ref{noise_amean_l100} show the 
eigenvalue distribution of the arithmetic mean of the $L$ non-Hermitian random matrix, 
for the noise only case.

\begin{figure}
 \centering
\includegraphics[width=3.4in]{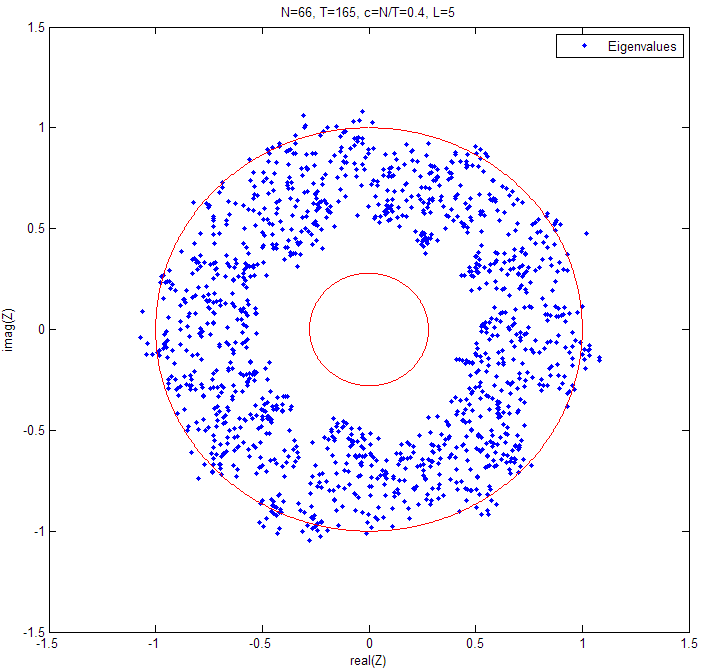}
\caption{The eigenvalue distribution for arithmetic mean of non-Hermitian random matrix, signal present, $L$=5.}
\label{signal_amean_l5}
\end{figure}

\begin{figure}
 \centering
\includegraphics[width=3.4in]{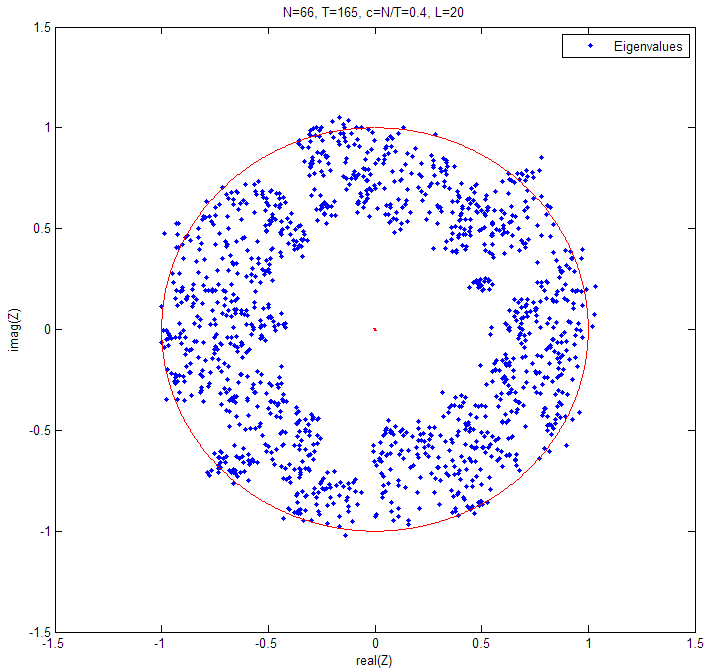}
\caption{The eigenvalue distribution for arithmetic mean of non-Hermitian random matrix, signal present, $L$=20.}
\label{signal_amean_l20}
\end{figure}

\begin{figure}
 \centering
\includegraphics[width=3.4in]{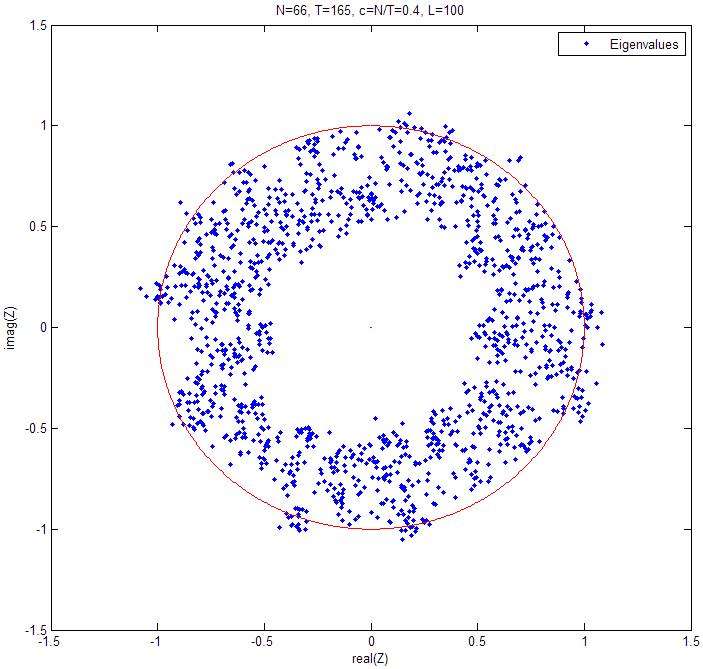}
\caption{The eigenvalue distribution for arithmetic mean of non-Hermitian random matrix, signal present, $L$=100.}
\label{signal_amean_l100}
\end{figure}

\textbf{Signal Present}: Fig.~\ref{signal_amean_l5} to Fig.~\ref{signal_amean_l100} show the 
eigenvalue distribution of the arithmetic mean of the $L$ non-Hermitian random matrix, 
for the signal present case.

\begin{figure}
 \centering
\includegraphics[width=3.4in]{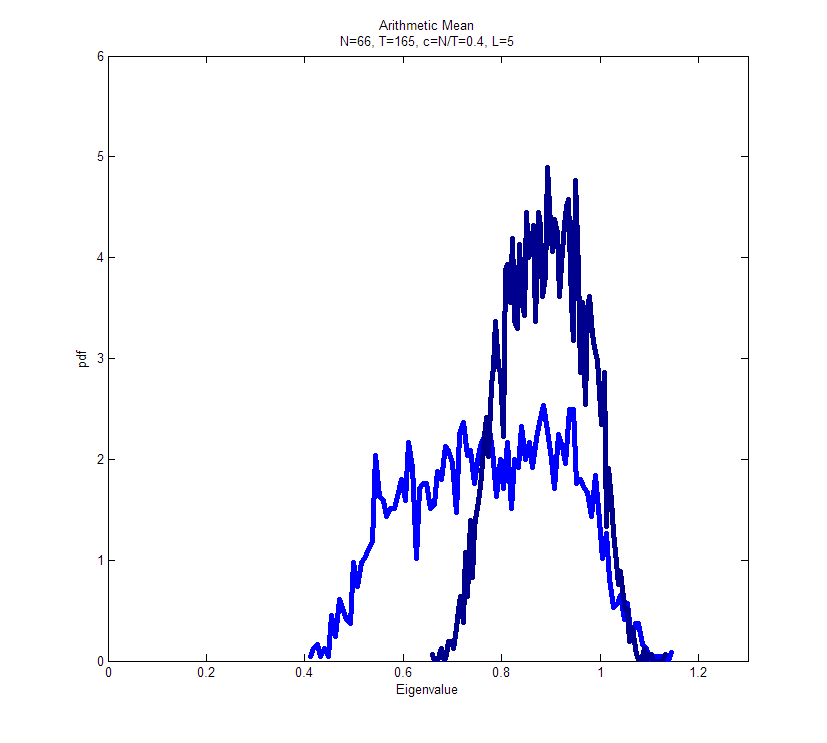}
\caption{Probability of eigenvalue for arithmetic mean of the non-Hermitian random matrix, both cases, with $L=5$.}
\label{pdf_ns_amean_L5}
\end{figure}

\begin{figure}
 \centering
\includegraphics[width=3.4in]{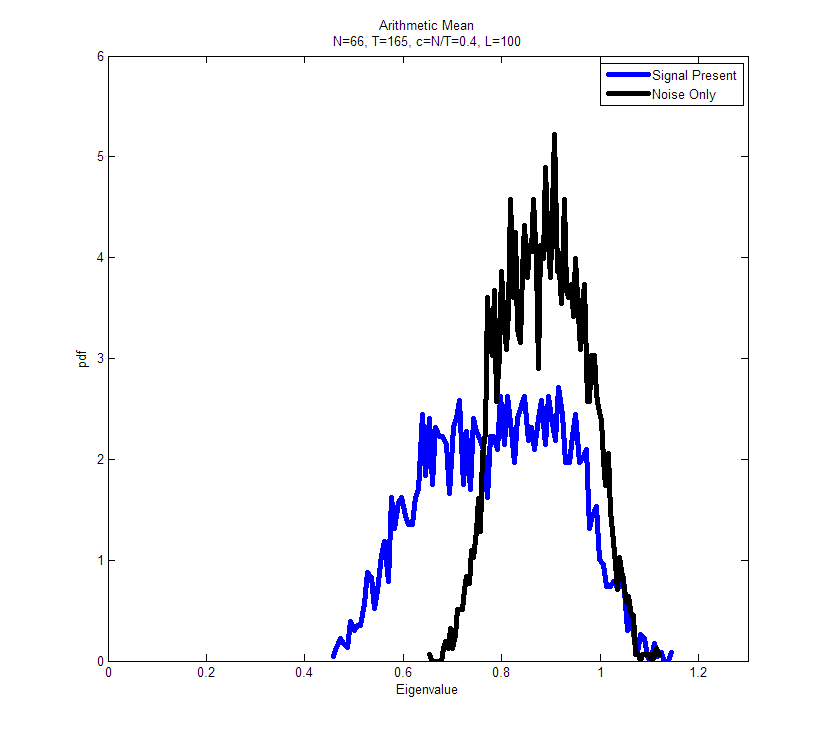}
\caption{Probability of eigenvalue for arithmetic mean of the non-Hermitian random matrix, both cases, with $L=100$.}
\label{pdf_ns_amean_L100}
\end{figure}

The corresponding PDFs of the eigenvalue absolute values of arithmetic mean are also shown Fig.~\ref{pdf_ns_amean_L5} and
Fig.~\ref{pdf_ns_amean_L100}.

From the visualized results of the eigenvalue distribution for Arithmetic mean model, we see 
\begin{itemize}
	\item The eigenvalue distribution for either noise-only and signal-present is following a similar ring law.
	\item The width of the ring, for signal-present, is larger than that for noise-only.
	\item We cannot get extra benefit by increasing $L$, as the width of the ring is not impacted by $L$.
\end{itemize} 

\subsection{Product of Random Ginibre Matrices}
We study the product of $k$ independent random square Ginibre matrices, ${\rm{Z}} = \prod\limits_1^k {{{\rm{G}}_i}} $. When the random Ginibre martices,
${{\rm{G}}_i}$, are square, the eigenvalues of ${\rm{Z}}{{\rm{Z}}^H}$ have asymptotic distribution ${\rho ^{\left( k \right)}}\left( x \right)$ in 
the large matrix limit. In terms of free probability theory, it is the free multiplicative convolution product of $k$ copies of the Marchenko-Pastur
distribution. In this model, we applied the space distributed approach to for the random matrix.

For $k = 2$, the spectral density is explicitly given by
\begin{equation}
{\rho ^{\left( 2 \right)}}\left( x \right) = \frac{{{2^{1/3}}\sqrt 3 }}{{12\pi }}\frac{{\left[ {{2^{1/3}}{{\left( {27 + 3\sqrt {81 - 12x} } \right)}^{2/3}} - 6{x^{1/3}}} \right]}}{{{x^{2/3}}{{\left( {27 + 3\sqrt {81 - 12x} } \right)}^{1/3}}}}
\end{equation}
where $x \in \left[ {0,{{27} \mathord{\left/
 {\vphantom {{27} 4}} \right.
 \kern-\nulldelimiterspace} 4}} \right]$.
For general $k$, the explicit form of the distribution is a superposition of hyper-geometric function of the type ${}_k{F_{k - 1}}$
%\begin{equation}
\begin{multline}
{\rho ^{\left( k \right)}}\left( x \right) = \sum\limits_{i = 1}^k {{\Lambda _{i,k}}{x^{\frac{i}{{k + 1}} - 1}}}  \cdot  \\
    {}_k{F_{k - 1}}\left( {\left[ {\left\{ {{a_j}} \right\}_{j = 1}^k} \right];\left[ {\left\{ {{b_j}} \right\}_{j = 1}^{i - 1},\left\{ {{b_j}} \right\}_{j = i + 1}^k} \right];\frac{{{k^k}}}{{{{\left( {k + 1} \right)}^{k + 1}}}}x} \right)
\end{multline}
%\end{equation}
where ${a_j} = 1 - \frac{{1 + j}}{k} + \frac{i}{{k + 1}},\;\;{b_j} = 1 + \frac{{i - j}}{{k + 1}},$ and 
%\begin{equation}
\begin{multline}
{\Lambda _{i,k}} = \frac{1}{{{k^{3/2}}}}\sqrt {\frac{{k + 1}}{{2\pi }}} {\left( {\frac{{{k^{k/\left( {k + 1} \right)}}}}{{k + 1}}} \right)^i}
 \cdot  \\ 
 \frac{{\left[ {\prod\limits_{j = 1}^{i - 1} {\Gamma \left( {\frac{{j - i}}{{k + 1}}} \right)} } \right]\left[ {\prod\limits_{j = k + 1}^k {\Gamma \left( {\frac{{j - i}}{{k + 1}}} \right)} } \right]}}{{\prod\limits_{j = 1}^k {\Gamma \left( {\frac{{j + 1}}{k} - \frac{i}{{k + 1}}} \right)} }}
\end{multline}
%\end{equation}
where ${}_p{F_q}\left( {\left[ {\left\{ {{a_j}} \right\}_{j = 1}^p} \right];\left[ {\left\{ {{b_j}} \right\}_{j = 1}^q} \right];x} \right)$ stands
for the hypergeometric function of the type ${}_p{F_q}$.

\begin{figure}
 \centering
\includegraphics[width=3.4in]{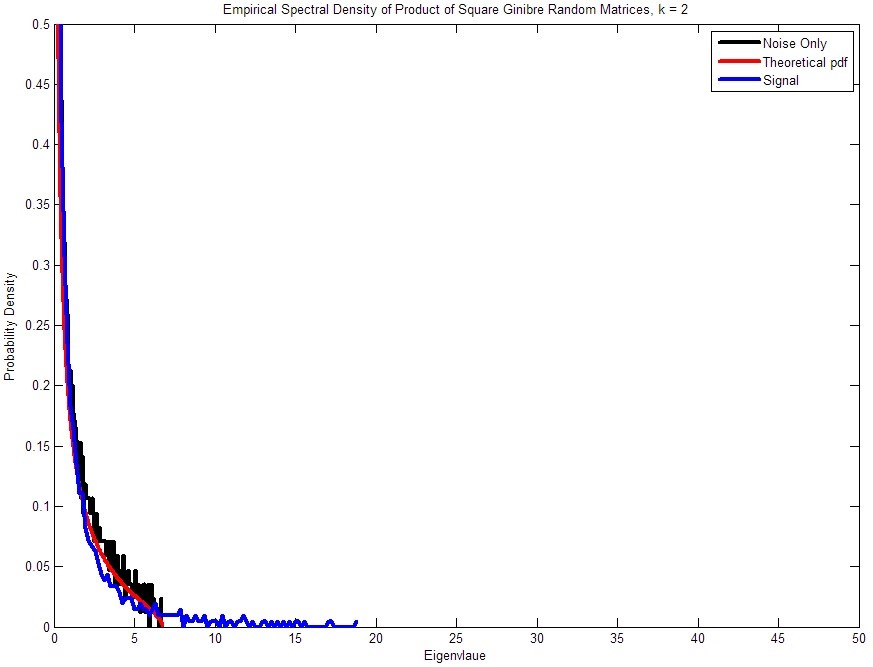}
\caption{Spectral density of eigenvalues for product of square Random Ginibre Matrices, k=2}
\label{pdf_ginibre_k2_noise}
\end{figure}

\begin{figure}
 \centering
\includegraphics[width=3.4in]{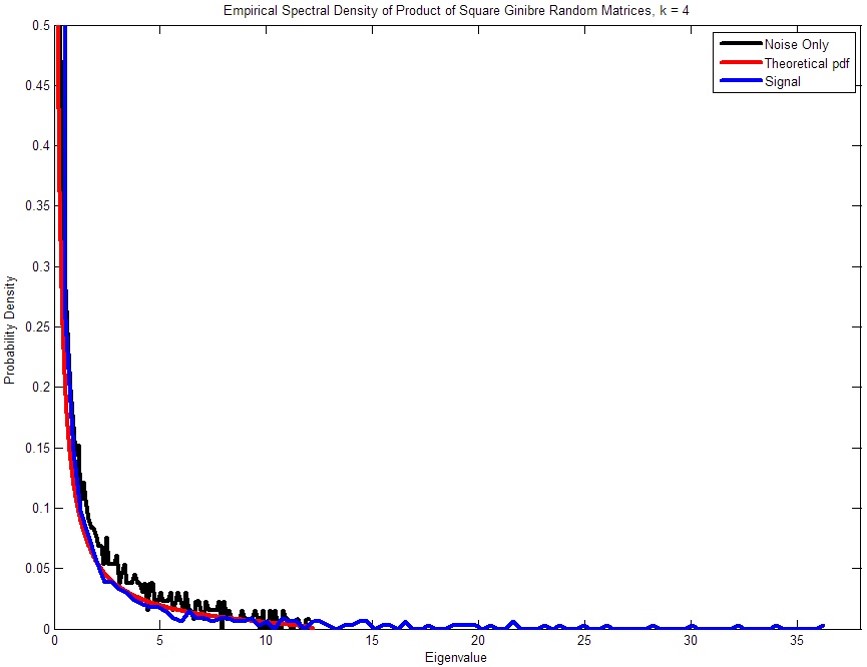}
\caption{Spectral density of eigenvalues for product of square Random Ginibre Matrices, k=4}
\label{pdf_ginibre_k4_noise}
\end{figure}

\begin{figure}
 \centering
\includegraphics[width=3.4in]{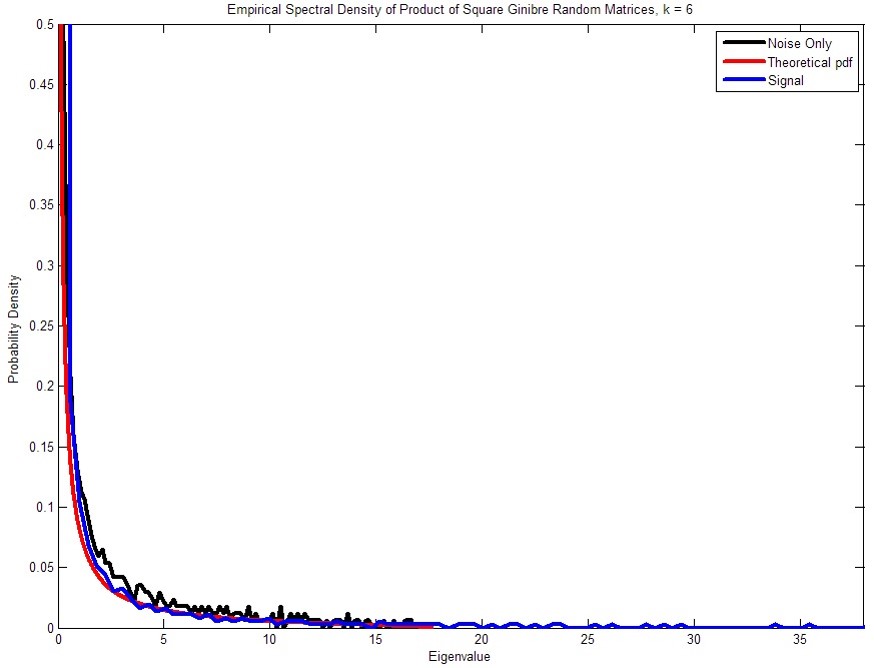}
\caption{Spectral density of eigenvalues for product of square Random Ginibre Matrices, k=6}
\label{pdf_ginibre_k6_noise}
\end{figure}

From the noise data captured by $k$ USRP sensors, we obtained the histogram for the spectral density of the product of the Ginibre random matrices. 
Fig.~\ref{pdf_ginibre_k2_noise} to Fig.~\ref{pdf_ginibre_k6_noise} show that the histograms match the theoretical pdf well, for different $k$.

\subsection{Summary of Theoretical Validation by Experimental Data}

We applied variant data models on the massive data collected by our massive MIMO testbed. Firstly, we found that the theoretical eigenvalue distribution (if exists)  can be validated by the experimental data for noise-only case. The random matrix based big data analytic model is successfully connected to the experiment. Secondly, the signal-present case can be differentiated from the noise-only case by applying the same data model. This result reveals the potential usage of the random-matrix based data model in signal detection, although the future work on the performance analysis is needed. 

\section{Initial Applications of Massive MIMO Testbed as Big Data System}
Besides signal detection, we demonstrated two applications based on the massive data analytic through the random-matrix method. The theoretical model in section ~\ref{proNH} is used, i.e., we mainly apply the product of non-Hermitian random matrices on the collected mobile data to investigate the corresponding eigenvalue distribution. Our aim is to make sense of massive data to find the hidden correlation between the random-matrix-based statistics and the information. Once correlations between “causes” and “effects” are empirically established, one can start devising theoretical models to understand the mechanisms underlying such correlations, and use these models for prediction purposes \cite{qiu2014smart}. 

\subsection{Mobile User Mobility Data }
In a typical scenario where the mobile user is communicating with the massive MIMO base station while moving, the uplink
waveform data received at each receiving antenna are collected. We applied the product of Hermitian random matrices to the data to observe the relationship between the eigenvalue distribution and the behavior of the moving mobile user. We are using the data from 10 antennas associated with 10 USRP receivers. Another USRP placed on a cart acts as the mobile user, which moves on the hallway of the 4th floor of the Clement Hall at Tennessee Technological University. The base station with up to 70 USRPs is on the same floor. The experiment results show that the moving speed of the mobile user is directly associated with the inner circle of the eigenvlaue distribution for the product of the Hermitian random matrices.

The experiments include five cases with different the moving speeds.
\paragraph{Case 1} \textbf{The Mobile User Stands in a Certain Place without Moving}

\begin{figure}[h!]
 \centering
\includegraphics[width=3.4in]{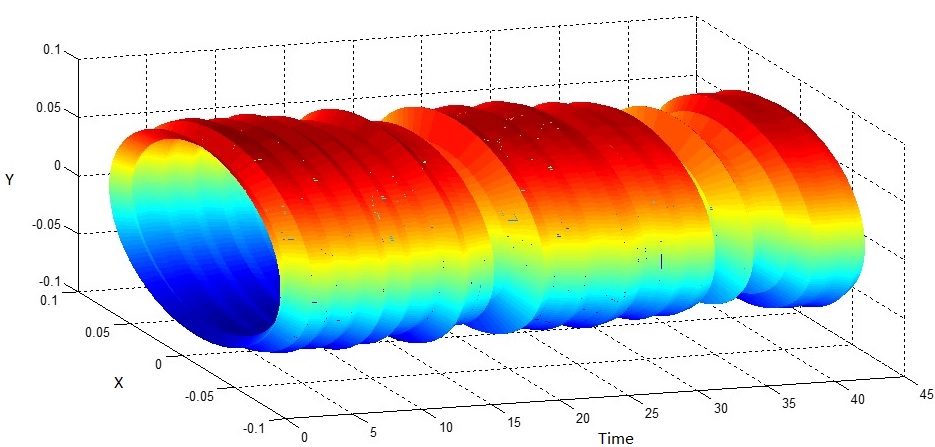}
\caption{Ring law inner radius changing with time for moving mobile user, case 1.}
\label{inner_nomove}
\end{figure}

In this case, the mobile user has zero speed. What we observed in Figure~\ref{inner_nomove} is that the inner radius
of the circle is almost not changing. The average inner radius is a little less than 0.05 for the whole procedure.

\paragraph{Case 2} \textbf{The Mobile User Moves at a Nearly Constant Walking Speed}

\begin{figure}[h]
 \centering
\includegraphics[width=3.4in]{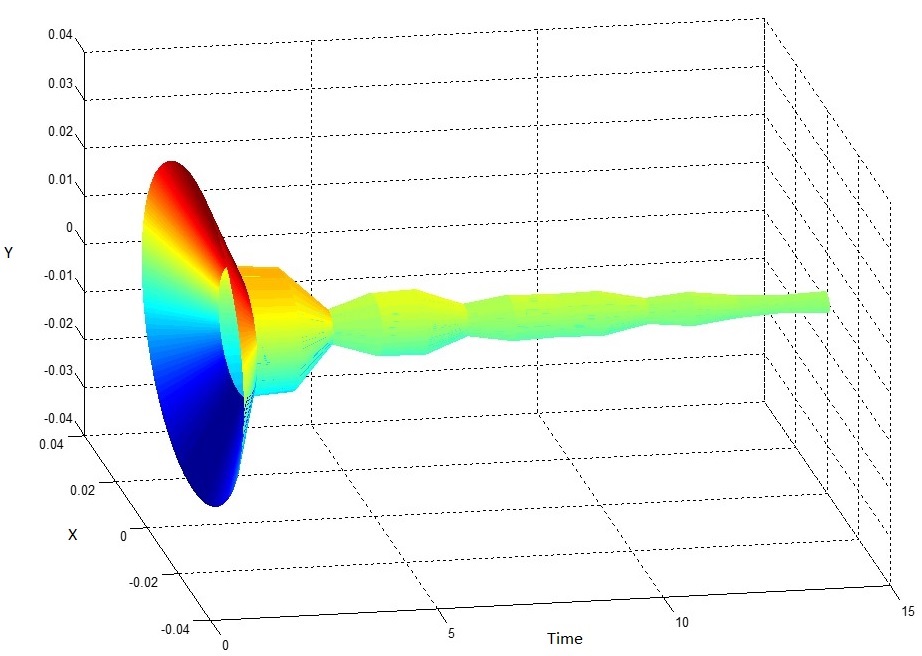}
\caption{Ring law inner radius changing with time for moving mobile user, case 2.}
\label{inner_walk}
\end{figure}

In this case, the mobile user moves along a straight line at a nearly constant walking speed from a distant point to a point near the base station. Figure~\ref{inner_walk} shows the change of the inner radius of the circle law with time. The moving mobile user is actually on a cart pushed by a man. We see the inner radius is much bigger at the beginning when the cart is accelerating from almost motionless to a walking speed than the rest of the time. During the moving stage, the inner radius is much smaller and very stable at around 0.005.

\paragraph{Case 3} \textbf{The Mobile User Moves at a Very slow speed}

\begin{figure}[h]
 \centering
\includegraphics[width=3.4in]{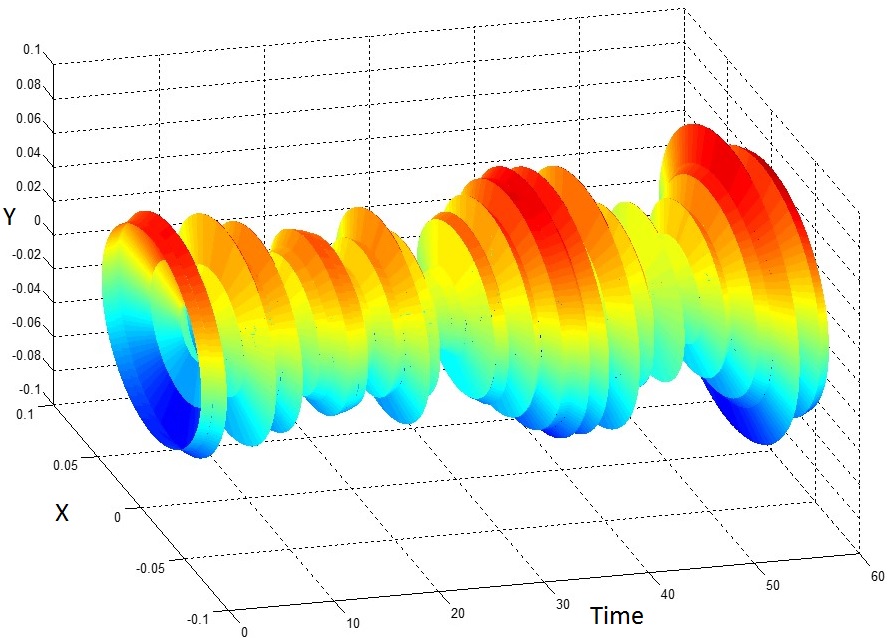}
\caption{Ring law inner radius changing with time for moving mobile user, case 3.}
\label{inner_slow}
\end{figure}

In this case, we move the mobile user at a very slow speed that is much smaller than walking speed. We see in Figure~\ref{inner_slow} that the inner radius is mostly vacillating between 0.02 and 0.05. This value is much smaller than that of the stationary case, but bigger than the walking-speed case. 

\paragraph{Case 4} \textbf{The Mobile User Moves at Varying speed: Half the Time walking, Half the Time at a Very Slow Speed.}

\begin{figure}[h]
 \centering
\includegraphics[width=3.4in]{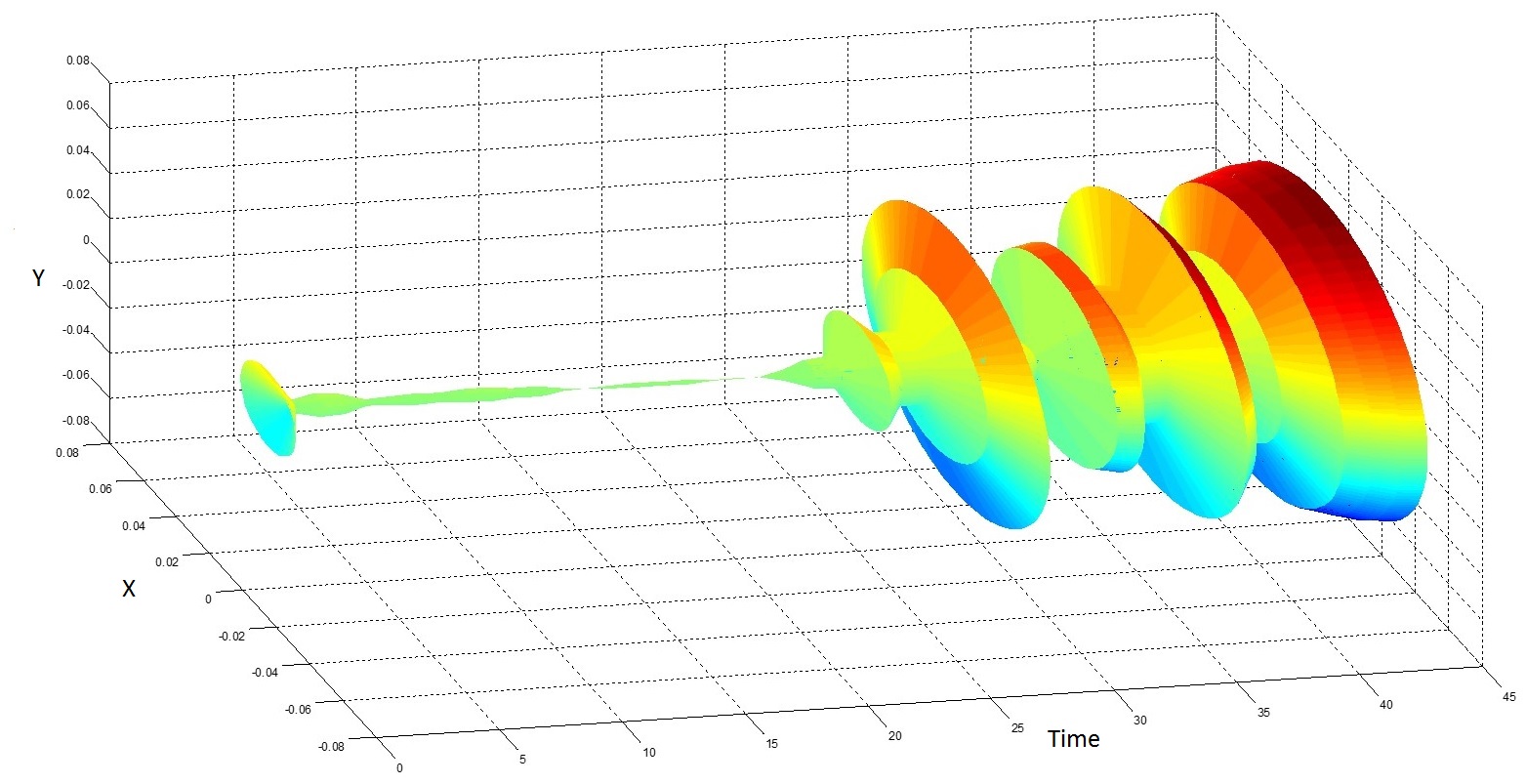}
\caption{Ring law inner radius changing with time for moving mobile user, case 4.}
\label{inner_vary1}
\end{figure}

In this case, we try to observe the difference for the impacts from different moving speeds on the inner radius in one figure.
Figure~\ref{inner_vary1} shows that the radius in the first half is much smaller than that in the second half. Correspondingly, the moving speed in the first half is much higher than the latter half.

\paragraph{Case 5} \textbf{The Mobile User Moves at Varying Speed: Half the Distance Walking, Half the Distance at a Very Slow Speed.}

\begin{figure}[h]
 \centering
\includegraphics[width=3.4in]{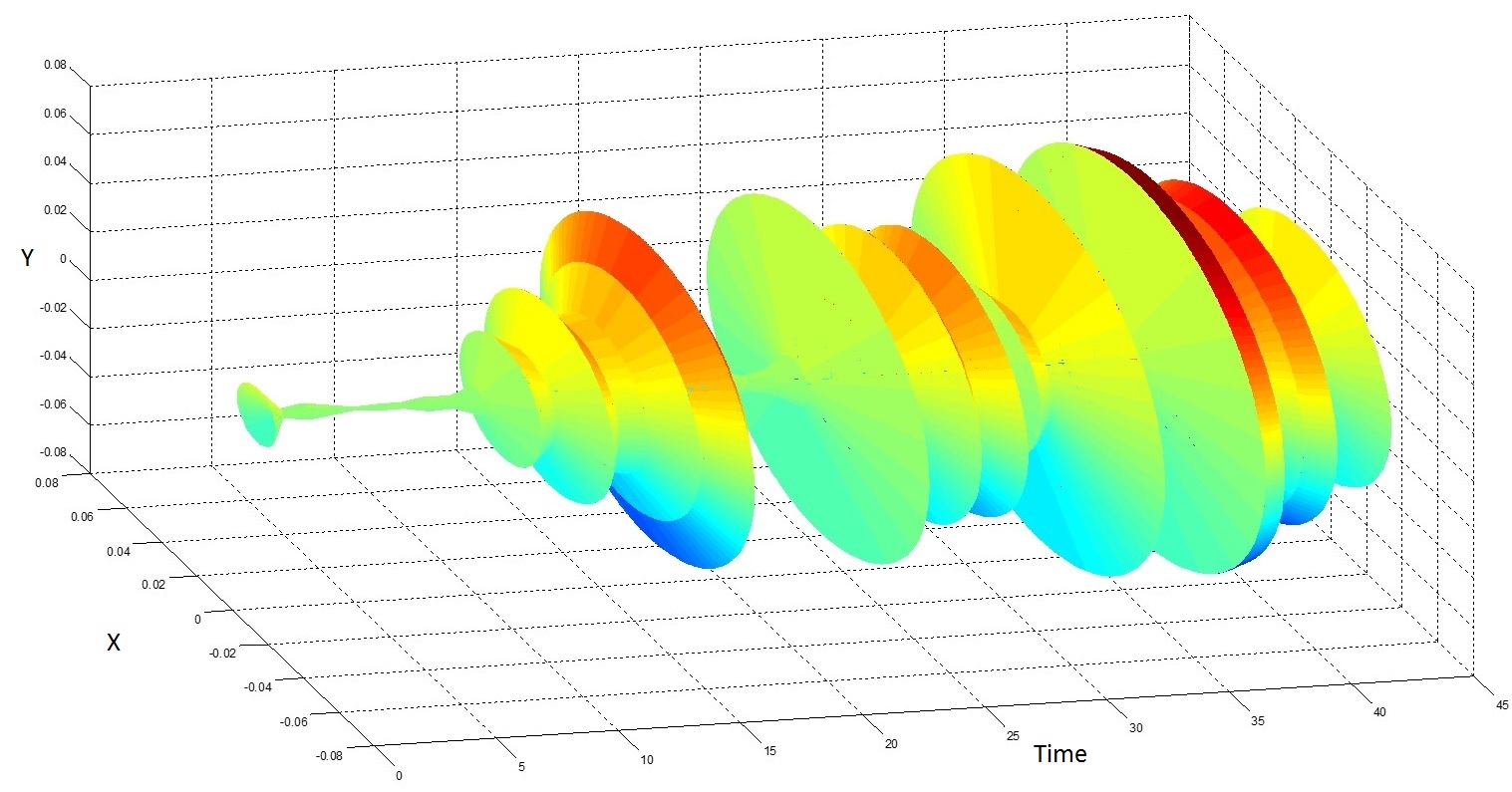}
\caption{Ring law inner radius changing with time for moving mobile user, case 5.}
\label{inner_vary2}
\end{figure}

Similar to case 4, the impacts from different speeds are observed in the figure. A higher moving speed brings a smaller inner radius of the eigenvalue distribution. Because the walking speed part has equal distance with the slow speed part, the occupied time of the former is smaller than the later part, just as shown in Figure~\ref{inner_vary2}.

%\paragraph{Case 6} The mobile user moves in a rectangular with almost constant walking speed.
%
%\begin{figure}[h]
 %\centering
%\includegraphics[width=3.2in]{inner_case6.jpg}
%\caption{.}
%\label{inner_rect}
%\end{figure}
All the above cases reveal a common observation that the faster the mobile user moves, the smaller the inner radius of ring law is. From the big data point of view, we can get insight that a massive MIMO based station can use the inner radius of the ring law to estimate the moving status of the mobile user. As we know, basically more correlation in the signal brings a smaller inner radius of the ring law. Thus, this result is reasonable, as the faster speed of the mobile user causes more Doppler effect to the random signal received in the massive MIMO base station, i.e., more correlation detected by the product of the Hermitain random matrices.

\subsection{Correlation Residing in Source Signal}

Besides the correlation introduced by the moving environment, as in the above experiment, the correlation residing in the transmitting signal also has a significant impact on the eigenvalue distribution of the random matrix. Actually, in the section on theoretical model validation, we only compared the cases of noise-only and signal-present. The correlation within the signal creates the derivation of the eigenvalue distribution. In this section, we intentionally adjust the auto-correlation level of the generated signal that is transmitted by the mobile user. The corresponding effect on the inner radius of the ring law is also investigated by analyzing the collected data from antennas at the massive MIMO base station.

%\subsubsection{Adjustable auto-correlation in the generated signal.}

\begin{figure}[h]
 \centering
\includegraphics[width=3.4in]{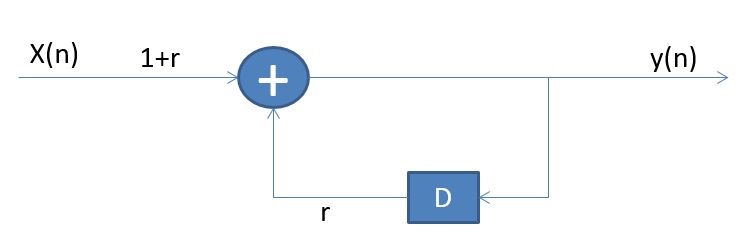}
\caption{Auto-regression filter used to generate the signal with adjustable autocorelation.}
\label{signal_gen}
\end{figure}

We generate the output signal following Eq.~\ref{sig_gen}:

\begin{equation}
y\left ( n \right )=\left ( 1+r \right )x\left ( n \right )+ry\left ( n-1 \right )
\label{sig_gen}
\end{equation}

which can also be represented by Figure~\ref{signal_gen}.
In the experiment, $x\left ( n \right )$ is set as Gaussian white noise.

\begin{figure}[h]
 \centering
\includegraphics[width=3.4in]{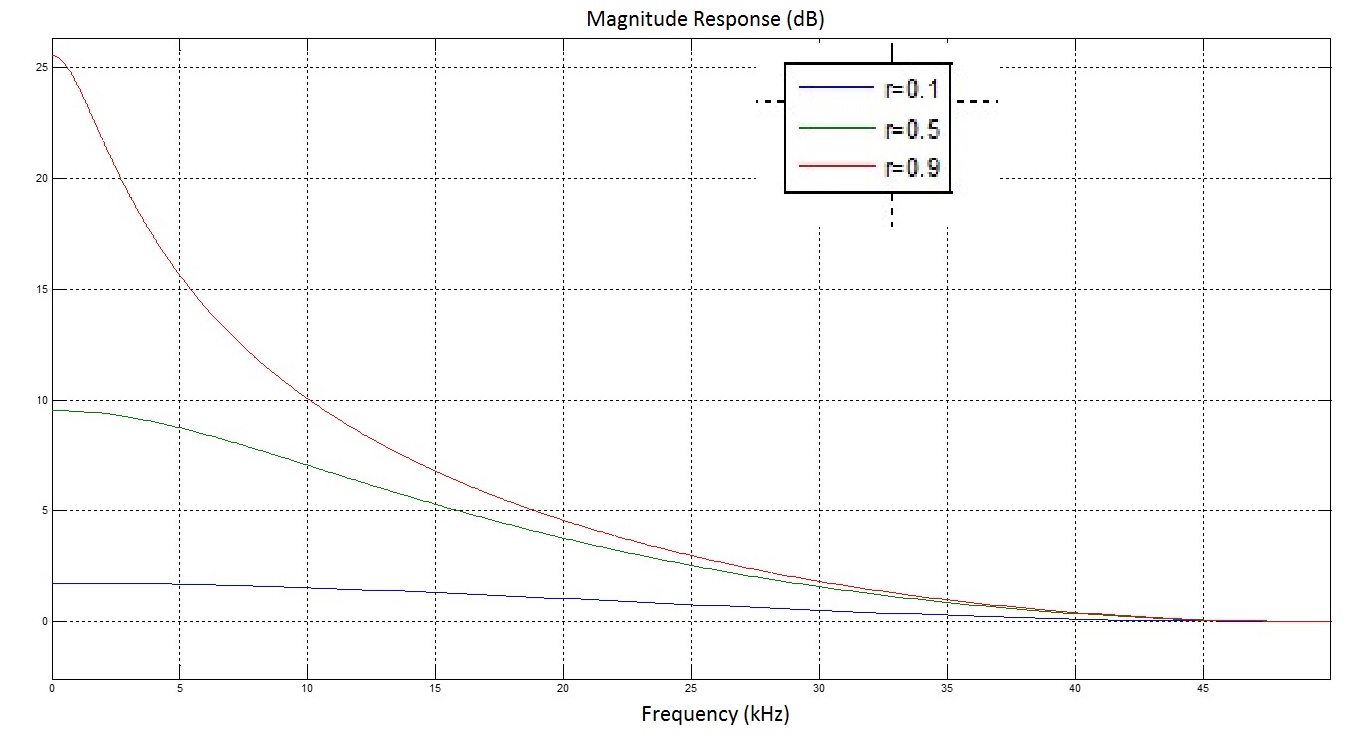}
\caption{Bigger $r$ leads to sharper frequency response of the AR filter for signal generator.}
\label{arfilter}
\end{figure}

\begin{figure}[h]
 \centering
\includegraphics[width=3.4in]{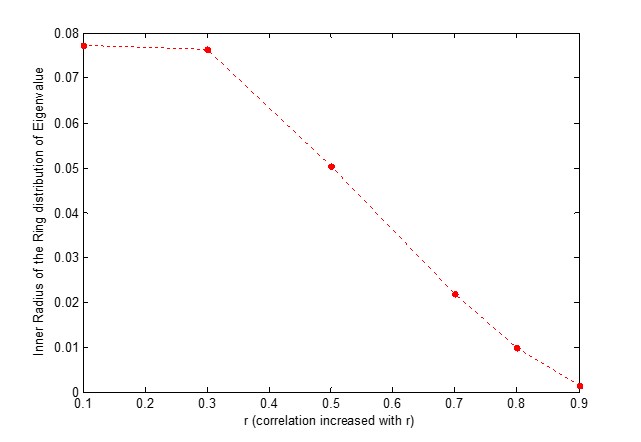}
\caption{Inner radius of ring law changes with the $r$, bigger $r$ leads smaller inner radius.}
\label{radius_r}
\end{figure}

Essential to this signal generator is an auto-regression filter in which the parameter $r$ is used to control the frequency response as shown in Figure~\ref{arfilter} 
A bigger $r$ leads a sharper frequency response that introduces more correlation within the transmitted signal. Thus, we can see that the inner radius of the ring law observed at the massive MIMO base station is as in Figure~\ref{radius_r}.
\subsection{Insights from Applications}

Both the applications bring us insights that the correlation residing in the signal can be matched to certain events in the network. In the network under our monitoring, such correlations can be detected and measured by our random-matrix-based data analysis method and finally be used to visualize the real event, such as the mobile user moving, fluctuation of the source signal correlation. This is a typical big data approach. The massive MIMO system is not only a communications system but also an expanded data science platform. We make sense of data by storing and processing the massive waveform data. Information will not be discarded, thus the energy of every bit/sample can be utilized as possible as we can. To our best knowledge, it is the first time, by concrete experiments, to reveal the value of the 5G massive MIMO as a big data system. We believe that more applications emerge in the future.

\section{Conclusion}
The paper gives a first account for the 70-node testbed that takes TTU four years to develop. Rather than focusing on the details of the testbed hardware development, we use the testbed as a platform to collect massive datasets. The motivated application of this paper is massive MIMO. First, by using our initial experimental data, we find that large random matrices are natural models for the data arising from this tested. Second, the recently developed non-Hermitian free probability theory makes the theoretical predictions very accurately, compared with our experimental results. This observation may be central to our paper. Third, the visualization of the datasets are provided by the empirical eigenvalue distributions on the complex plane. Anomaly detection can be obtained through visualization. Fourth, when no visualization is required, we can formulate spectrum sensing or network monitoring in terms of matrix hypothesis testing. This formulation is relatively new in our problem at hand for massive MIMO. To our best knowledge, our work may be the first time. A new algorithm is proposed for distributed data across a number of servers.

At this moment of writing~\cite{Qiu2016MassiveDataAnalysis}, we feel that both theoretical understanding and experimental work allows for extension to other applications. First, thousands of vehicles need be connected. Due to mobility, streaming data that are spatially distributed across $N=1,000$ becomes essential. We have dealt with hypothesis testing problem. How do we reduce the data size while retaining the statistical information in the data? Sketching~\cite{woodruff2014sketching} is essential~\cite{Qiu2016MassiveDataAnalysis}.  Second, the testbed allows for the study of data analytical tools that will find applications in large-scale power grid, or Smart Grid~\cite{QiuAntonik2014Wiley}. For example, the empirical eigenvalue distribution of large random matrices is used for power grid in~\cite{he2015big}.

\section*{Acknowledgment}
This work is funded by the National Science Foundation through three grants (ECCS-0901420, ECCS-0821658, and CNS-1247778), and the Office of Naval Research through two grants (N00010-10-1-0810 and N00014-11-1-0006). 
%\cite{joshi2010adaptive,ye2008energy,paliwal1988estimation}

% Can use something like this to put references on a page
% by themselves when using endfloat and the captionsoff option.
\ifCLASSOPTIONcaptionsoff
  \newpage
\fi

% trigger a \newpage just before the given reference
% number - used to balance the columns on the last page
% adjust value as needed - may need to be readjusted if
% the document is modified later
%\IEEEtriggeratref{8}
% The "triggered" command can be changed if desired:
%\IEEEtriggercmd{\enlargethispage{-5in}}

% references section

	%\newpage \setcounter{page}{1}

    \bibliographystyle{ieeetr}   

 \begin {small}
 
 %\bibliography{Bible/bibmimo}

\bibliography{Bible/bibmimo,Bible/5GWirelessSystem,Bible/Big_Data,Bible/Compressed_Sensing,Bible/Smart_Grid,Bible/Graph_Complex_Network,Bible/Machine_Learning,Bible/Convex_Optimization,Bible/LowRankMatrixRecovery,Bible/Concentration_of_Measure,Bible/ClassicalMatrixInequalities,Bible/CompeltelyPositiveMaps,Bible/QuantumChannel,Bible/QuantumHypothesisTesting,Bible/TraceInequalities,Bible/QuantumInformation,Bible/Matrix_Inequality,Bible/RandomMatrixTheory,Bible/Fractional_Integration_bib,Bible/UWB_bib,Bible/Qiu_Group_bib,Bible/LTI_Comm_Theory_bib,Bible/Software_Defined_Radio_bib,Bible/Time_Reversal_bib,Bible/MIMO_bib,Bible/Radar_Waveform_Optim_bib,Bible/Information_Theory_bib,Bible/Compressed_Sensing_Theory,Bible/Compressed_Sensing_UWB,Bible/MISC_bib,Bible/CS_Applications,Bible/CS_Data_Stream_Algorithms,Bible/CS_Extensions,Bible/CS_Foundations_Connections,Bible/CS_Multisensor_Distributed,Bible/CS_Recovery_Algorithms,Bible/CognitiveRadio/Cognitive_Radio_bib,Bible/CognitiveRadio/CognitiveRadio2008_bib,Bible/CognitiveRadio/DySpan2007,Bible/CognitiveRadio/DySpan2005,Bible/CognitiveRadio/Gardner,Bible/CognitiveRadio/jsac200703,Bible/CognitiveRadio/jsac200801,Bible/CognitiveRadio/sensingDTV,Bible/CognitiveRadio/ucberkeley,Bible/CognitiveRadio/UWBCognitiveRadio_bib,Bible/CognitiveRadio/IEEE_JSSP_2008,Bible/CognitiveRadio/Bayesian_Network_Cognitive_Radio,Bible/CognitiveRadio/Exploiting_Historical_Spectrum,Bible/CognitiveRadio/Duke_Carin,Bible/CognitiveRadio/LiHu_upper,Bible/CognitiveRadio/AFRLref,Bible/CognitiveRadio/SVM,Bible/CognitiveRadio/NSF_ECCS_0821658,Bible/CognitiveRadio/SmartGrid}

\end {small}
\end{document}